\documentclass[aps,10pt,prd,notitlepage]{revtex4-1}
\usepackage[utf8]{inputenc}
\usepackage{amsmath,amssymb,bm,epsfig}
\usepackage{color}
\usepackage{hyperref} 
\usepackage{natbib}
\usepackage{ulem}
\usepackage{graphicx}
\usepackage{xcolor}
\usepackage{pifont}
\usepackage{amsfonts}
\usepackage{dsfont,mathrsfs}
\usepackage{cancel}
\usepackage{bm}
\usepackage{bigints}
\newcommand{\nn}{\nonumber}

\newcommand{\ket}[1]{\left|#1\right\rangle}

\newcommand{\outerproduct}[2]{\left|#1\right\rangle\left\langle#2\right|}
\newcommand{\expectationvalue}[3]{\left\langle#1\left|#2\right|#3\right\rangle}
\newcommand{\ensembleaverage}[1]{\left\langle#1\right\rangle}

\newcommand{\MB}[1]{\left|#1\right|}
\newcommand{\FB}[1]{\left(#1\right)}
\newcommand{\SB}[1]{\left\{#1\right\}}
\newcommand{\TB}[1]{\left[#1\right]}

\newcommand{\scrL}{\mathscr{L}}

\newcommand{\sign}[1]{\text{sign}\left(#1\right)}

\newcommand{\munu}{{\mu\nu}}
\newcommand{\numu}{{\nu\mu}}
\newcommand{\alphabeta}{{\alpha\beta}}

\newcommand{\IM}{\text{Im}}
\newcommand{\RE}{\text{Re}}
\newcommand{\Tr}{\text{Tr}}

\begin{document}
\title{Electromagnetic spectral function and dilepton rate in a hot magnetized QCD medium}
\author{Snigdha Ghosh}
\email{snigdha.physics@gmail.com, snigdha.ghosh@iitgn.ac.in}
\author{Vinod Chandra}
\email{vchandra@iitgn.ac.in}
\affiliation{Indian Institute of Technology Gandhinagar, Palaj, Gandhinagar 382355, Gujarat, India}

\begin{abstract}
The dilepton production rate in hot QCD medium is studied within a effective description of the medium in the presence of magnetic field. This could be done by obtaining the one-loop self energy of photon due to the effective (quasi-) quark loop at finite temperature under an arbitrary external magnetic field while employing the real time formalism of Thermal Field Theory. The effective quarks and gluons encode hot QCD medium effective in terms of their respective effective fugacities. The magnetic field enters in the form of landau level quantization, in the matter sector (quarks, antiquarks). The full Schwinger proper time propagator including all the Landau levels is considered for the quasi quarks while calculating the photon self energy. The electromagnetic Debye screening (in terms of the self-energy) has seen to be influenced both by the hot QCD medium effects and magnetic field. Analogous results are also obtained from the semi classical transport theory. The imaginary part of the photon self energy function is obtained from the discontinuities of the self energy across the Unitary cuts which are also present at zero magnetic field and the Landau cuts which are purely due to the magnetic field. The dilepton production rate is then obtained in terms of  the product of electromagnetic spectral functions due to quark loop and lepton loop. The modifications of both the quarks/antiquarks as well as leptons in presence of an arbitrary external magnetic field have been considered in the formalism. Significant enhancement of the low invariant mass dileptons due the appearance of the Landau cuts in the electromagnetic spectral function at finite external magnetic field has been observed. A substantial enhancement of dilepton rate is also found when the EOS effects are considered through the effective quarks/antiquraks.
\end{abstract}

\maketitle
%
\section{Introduction}\label{sec.intro}
The study of the  nuclear matter under extreme conditions of temperature and/or density has been  a subject of intense investigation over past few decades. Heavy Ion Collision (HIC) experiments in Relativistic Heavy Ion Collider (RHIC) and Large Hadron Collider (LHC) are expected to produce such state in the laboratory where effectively  free quarks and gluons become nearly thermalized over nuclear volume scale. This form of the nuclear matter is commonly termed as Quark-Gluon-Plasma (QGP)  which is more like a near perfect~\cite{Adams:2005dq, Adcox:2004mh, Back:2004je, Arsene:2004fa} and most vortical fluid~\cite{STAR:2017ckg} that could be created in the experiments. The experimental observations strongly indicate towards the strongly interacting nature of the QGP. Recent studies have further revealed that, in a non-central or asymmetric HIC, extremely high magnetic fields are created~\cite{Skokov:2009qp}. The magnitude of the magnetic field is comparable to the typical QCD energy scale ($eB\sim \Lambda_\text{QCD}^2$) and thus it could non-trivially  affect bulk as well microscopic properties of the QGP.  In fact the strength of the magnetic field turned out to larger that  fields  present in the interior of a magnetar~\cite{Duncan:1992hi} though in the very early stages of the HIC.

Next, the study of this strongly interacting matter at high temperature and density along with external magnetic field has gained extreme momentum  over the last decade~\cite{Kharzeev:2012ph} because of many exotic effects like Chiral Magnetic Effect(CME), Magnetic Catalysis (MC), Inverse Magnetic Catalysis (IMC), Chiral Vortical Effect (CVE), superconductivity of the vacuum {\it etc}.~\cite{Kharzeev:2007tn, Kharzeev:2007jp, Fukushima:2008xe, Gusynin:1995nb, Gusynin:1999pq, Bali:2011qj, Chernodub:2010qx, Chernodub:2012tf} that can happen when the hot and dense QCD matter is placed under high external magnetic field.  Notably,  these studies are also important in the context of astrophysics and cosmology~\cite{Duncan:1992hi, Ferrer:2005vd, Ferrer:2006vw, Ferrer:2007iw, Fukushima:2007fc, Feng:2009vt, Fayazbakhsh:2010gc, Fayazbakhsh:2010bh}.

The QGP formed in a HIC experiment is a `transient' state, which exists for a very short time ($\sim$ few fm/c) implying that it can not be observed directly (the direct detection is also prohibited by the color confinement). There are various indirect probes and observables~\cite{Wong:1995jf} that are used to extract the properties of this hot and dense medium such as electromagnetic probes (photon and dileptons)~\cite{Arnold:2001ms, Aurenche:1999tq, Aurenche:2000gf, Chatterjee:2009rs, Alam:1996fd, Alam:1999sc, Kajantie:1986dh, McLerran:1984ay, Rapp:1999ej, Weldon:1990iw}, heavy quarks~\cite{Rapp:2009my}, quarkonia~\cite{Matsui:1986dk}, jets~\cite{Wang:1991xy}, collective flow~\cite{Poskanzer:1998yz, Voloshin:2008dg, Ollitrault:1992bk, Kolb:2003dz, Adams:2005dq, Aamodt:2010pa} {\it etc}. Another important theoretical tool to probe the microscopic properties of the medium has been the  study of different $n$-point current-current correlation functions at finite temperature and density (the in-medium spectral functions of local currents). The transport quantities like shear and bulk viscosities, thermal conductivity {\it etc}. could be obtained from the spectral functions consisting of appropriate currents. In our context, the prime interest  is in the electromagnetic spectral function, which is obtained from the vector-vector current correlator. The correlator  can also be related to the Dilepton Production Rate (DPR) from the hot and dense magnetized QGP medium.  The dileptons are emitted from the entire space-time volume throughout the medium evolution. Because of the fact that, the dileptons interact only through the electromagnetic interaction, they have larger mean free paths. The dileptons  come out of the thermal medium soon after their production  without suffering more collisions. Thus the dileptons  carry the precise information of the thermodynamic state of the medium where they are produced. In the QGP medium, a quark interacts with an antiquark to produce a virtual photon which subsequently decays into a dilepton. In addition to this, there are other sources of dileptons in a HIC experiment such as  the interaction of charged hadrons with their antiparticles like $\pi^+\pi^-\rightarrow l^+l^-$ can  produce the dileptons. They can also be produced from the decays of hadron resonances ($\pi^0,\rho,\omega,J/\psi$ {\it etc}). The Drell-Yan process also gives significant contribution to the high invariant mass dilepton productions.

In order to calculate the electromagnetic spectral function and DPR, the most essential input is the local equilibrium distribution functions of the quarks/antiquarks that describes the interacting hot QCD/QGP medium. At this juncture, a recently developed  Effective fugacity Quasi Particle Model (EQPM)~\cite{Chandra:2011en} provides a systematic description of hot QCD medium in accordance with the realistic QCD Equation of State (EOS) in terms of  effective temperature dependent fugacities in the thermal distribution of gluons and quarks/antiquarks which encode all the effect of strong interaction. The temperature dependences of the effective fugacities could be obtained from the realistic hot QCD equations of state such as  recent lattice QCD EOS.

The photon polarization tensor under external magnetic field has been calculated earlier at zero temperature in Refs.~\cite{Hattori:2012je, Hattori:2012ny, Chao:2014wla, Tsai:1974ap} and also at finite temperature in Ref.~\cite{DOlivo:2002omk}. The DPR in presence of external magnetic field has been studied earlier in  Refs.~\cite{Tuchin:2012mf, Tuchin:2013bda, Sadooghi:2016jyf, Mamo:2013efa, Bandyopadhyay:2016fyd, Bandyopadhyay:2017raf}. In Refs.~\cite{Tuchin:2012mf, Tuchin:2013bda}, the authors have obtained the DPR from hot magnetized QGP in a phenomenological way including the effects of synchrotron radiation as well as quark-antiquark annihilation. Where as in~\cite{Sadooghi:2016jyf}, the authors have used the Ritus formalism to calculate the photon polarization tensor and DPR under external magnetic field. Recently the DPR in strong as well as in weak magnetic field approximation has been obtained in Refs.~\cite{Bandyopadhyay:2016fyd, Bandyopadhyay:2017raf}. In all of mentioned works, the authors have considered the ideal Fermi-Dirac distribution function for the quarks/antiquarks for the calculation of DPR. In contrast, in Ref.~\cite{Chandra:2015rdz}, the DPR has been obtained using the modified equilibrium distribution functions of quarks/antiquarks within the EQPM at zero external magnetic field where the effect of viscous modification are also included. This sets the motivation for the present work, where we aim for a systematic inclusion of magnetic field effects and realistic QGP equation of state in the field theoretical formalism.

In this work, we aim to calculate the photon polarization tensor with the \textit{effective quarks} as the loop particles (incorporating EQPM) in presence of arbitrary external magnetic field at finite temperature. The Real Time Formalism (RTF) of Thermal Field Theory (TFT) has been employed for the calculation of one-loop photon self energy at finite temperature. The full  Schwinger proper time propagator including all the Landau levels for the quasi quark propagation is considered in the calculation. No strong or weak field approximation has been made in the analysis as usually done in most of the works in the literature. The Debye screening mass  for the electromagnetic screening in the hot magnetized QCD medium is obtained  by taking static limit of the $00$ component of the photon polarization tensor ($\Pi^{\mu\nu}$) and it matches  with the  one  obtained from the semi-classical transport theory. Further, the imaginary part of the self energy is obtained from the discontinuities across the Unitary and Landau cuts. For physical time like momentum of the external photon,   the Unitary as well as the Landau cuts are observed. The Unitary cuts are also present at zero temperature and zero external magnetic field whereas the Landau cuts only appear at finite temperature and only when the loop particles have different masses. In our case, although the loop particles are both quarks having equal mass, yet the Landau cuts appear due to the external magnetic field. The appearance of the Landau cuts have correspondence to the physical processes like photon emission or absorption by a quark/antiquark in magnetized QGP which was forbidden in zero magnetic field case due to kinematic restrictions. The  analytic structure of the imaginary part of the self energy has also been analyzed and we found that the thresholds of the Unitary and Landau cuts get modified due to the presence of external magnetic field. Finally, the DPR is obtained  from the thermal QGP medium while expressing  in terms of the the electromagnetic spectral functions due to quark loop and lepton loop.

Note that there has been another method used in the literature for obtaining the leptonic part in DPR is by taking into account the spin sums over the leptonic spinor~\cite{Mallik:2016anp, Bandyopadhyay:2016fyd}. We have shown explicitly that these two different approaches converge to the same result. However, this particular form of the DPR in terms of spectral function due to quark and lepton loop enables us to introduce the external magnetic field through the modification of the quark and lepton propagator in terms of Schwinger proper time one where a knowledge of spin sum over leptonic spinor in presence of external magnetic field is not required.  We have found  significant impact on the low invariant mass dilepton yield due to external magnetic field. The DPR also shows significant enhancement when the QCD medium effects are considered via the EQPM.

The paper is organized as follows. In Sec.~\ref{sec.eqpm}, we have briefly discussed the EQPM. In Sec~\ref{sec.self.t}, the one-loop photon self energy due to quark loop is obtained at finite temperature in absence of external magnetic field. This is followed by the calculation of the photon polarization tensor under external magnetic field in addition to finite temperature in Sec.~\ref{sec.self.tb}. After that, we have discussed the Debye screening mass in Sec.~\ref{sec.debye} and the electromagnetic spectral functional along with its analytic structure in Sec.~\ref{sec.rho}. Sec.~\ref{sec.dlr.0} and \ref{sec.dlr.0} are devoted for the calculation of DPR at zero and finite external magnetic field respectively. The numerical results are provided in Sec~\ref{sec.numerical} and finally we summarize and conclude in Sec~\ref{sec.summary}. Some of the relevant calculational details are provided in the Appendices.


\section{The EQPM} \label{sec.eqpm}
The EQPM describes hot QCD medium effects that are encoded in the realistic hot QCD EOSs in terms of a grand canonical system of effective quarks, antiquarks and gluons. In this scheme, the hot QCD medium effects encoded in the realistic hot QCD EOSs (either computed from lattice QCD or HTL perturbation theory) could be introduced in terms of effective quark/antiquark and effective gluon distribution functions. In this work, the EQPM developed in Ref.~\cite{Chandra:2011en} and later generalized for the magnetized hot QCD medium in Ref.~\cite{Kurian:2017yxj} is utilized to include hot magnetized QCD medium effects. The EQPM maps the hot QCD EOSs in terms of effective quark-antiquarks and effective gluons with effective fugacities. The effective fugacities in the model are merely introduced to capture the hot QCD medium effects and could be understood in terms of non-trivial dispersion for effective quarks and gluons. They should not be confused with the presence of any conserved current. Their physical meaning is reflected in the modified dispersion of quasi particles and in terms of mean field contribution in the effective covariant kinetic theory~\cite{Mitra:2018akk}.

Notably, there are other effective quasi particle models to describe the hot QCD medium effects  with effective masses for quarks and gluons~\cite{Goloviznin:1992ws, Peshier:1995ty}, effective masses  with Polyakov loop~\cite{DElia:1997sdk, DElia:2002hkf, Castorina:2005wi, Castorina:2007qv}, NJL and PNJL based quasi-particle models~\cite{Dumitru:2001xa, Fukushima:2003fw, Ghosh:2006qh, Abuki:2009dt, Tsai:2008je}, and self-consistent and single parameter quasi-particle models~\cite{Bannur:2006hp, Bannur:2006js, Bannur:2006ww}. There are a few other recently proposed quasi-particle models based on the Gribov-Zwanziger (GZ) quantization~\cite{Su:2014rma, Florkowski:2015dmm, Florkowski:2015rua, Bandyopadhyay:2015wua}. All these models tried to interpret interacting hot QCD medium in terms of non-interacting/weakly interacting  quasi-particles with some effective parameters that capture all the medium effects.

The EQPM  describes the quasi-gluon and quasi quark/antiquark in terms of their respective distribution functions
\begin{eqnarray}\label{3}
\tilde{f}(|\vec{k}|)=\frac{1}{z^{-1}_g e^{\beta |\vec{k}|} - 1 } ~~\text{and}~~
f(\omega_k^f)=\frac{1}{z^{-1}_q e^{\beta \omega_k^f} + 1 } \label{eq.distribution.function}
\end{eqnarray}    
where $\omega_{k}^f=\sqrt{\vec{p}^{~2}+m_f^{2}}$ with $m_f$ being the mass of quarks/antiquark flavour $f$. The physical significance of the effective fugacities $z_{g,q}$ comes in the dispersion relations of gluons and quarks as 
\begin{equation}\label{4}
\tilde{\omega}_{k}^g=|\vec{k}|+T^{2}\frac{\partial}{\partial T} \ln(z_{g}) ~~\text{and}~~   \tilde{\omega}_{k}^f=\omega_k^{f}+T^{2}\frac{\partial}{\partial T} \ln(z_{q})~.
\end{equation}

Both $z_g$ and $z_q$ have complicated temperature dependence as discussed in Ref.~\cite{Jamal:2017dqs}. Here, we consider the EQPM description of the recent (2+1) flavor lattice QCD EOS~\cite{Bazavov:2009zn, Bazavov:2014pvz}. 
Notably, the magnitudes of  $z_g$ and $z_q$ are always less than unity. The asymptotic limit, $z_q = z_g = 1$  is equivalent to ideal EOS (Stefan-Boltzmann limit) for the QCD.  The fugacities , 
 $z_g=z_g(T) <1 $ and $z_q=z_q(T)<1$ have been constructed in such a way, that a grand canonical ensemble 
of the non-interacting quasi-partons leads to an effective EOS identical to the LQCD one. In this work we have presented the numerical results 
of all the quantities for two different cases separately: (i) $z_q = z_g = 1$ abbreviated as ``Ideal EOS" and (ii) $z_q = z_q(T)<1$, 
$z_g = z_g(T) <1$ abbreviated as ``LQCD EOS".  
The extension of EQPM in the magnetic field background involves the modification of dispersion relation by relativistic Landau levels as discussed in~\cite{Kurian:2017yxj}. Extended EQPM for hot magnetized QGP has further been employed to study the transport coefficients of medium~\cite{Kurian:2018qwb, Kurian:2018dbn} while employing the covariant kinetic theory developed in context of EQPM in Ref.~\cite{Mitra:2018akk}.

\section{Photon Self Energy at Finite Temperature}\label{sec.self.t}
The Lagrangian for $\gamma q\bar{q}$ electromagnetic interaction is given by
\begin{eqnarray}
\scrL_\text{int} = \sum_{f}^{}e_f\bar{q}_f\gamma^\mu q_fA_\mu 
\label{eq.lagrangian}
\end{eqnarray}
where, $q_f(\bar{q}_f)$ is the quark(antiquark) field, $A_\mu$ is the photon field and $e_f=\sqrt{4\pi\alpha}Q_f$ is the electric charge for quark flavour $f$. Here $Q_f=\frac{2}{3}$ for $f\equiv(u,c,t)$ and $Q_f=-\frac{1}{3}$ for $f\equiv(d,s,b)$. $\alpha=\frac{1}{137}$ is the QED fine structure constant. 
\begin{figure}[h]
	\begin{center}
		\includegraphics[angle=0, scale=0.5]{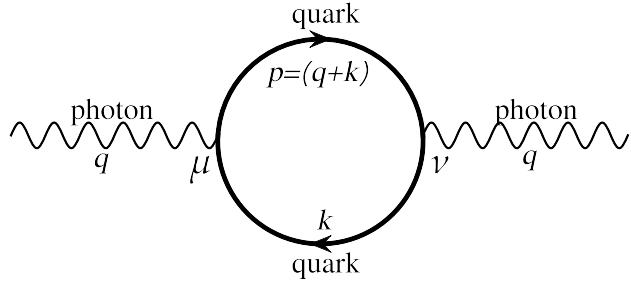}
	\end{center}
	\caption{Feynman diagram for one-loop photon self energy.}
	\label{fig.feynman}
\end{figure}
Using Eq.~(\ref{eq.lagrangian}), the one-loop vacuum self energy of photon can be written as (By applying Feynman rule to Fig.~\ref{fig.feynman}),
\begin{eqnarray}
\Pi_\text{vac}^\munu(q) = -\sum_{f}^{}e_f^2N_ci\int\frac{d^4k}{(2\pi)^4}\Tr\TB{\gamma^\nu S_F^f(p=q+k)\gamma^\mu S_F^f(k)}
\end{eqnarray}
where, $N_c$ is the number of colour and 
\begin{eqnarray}
S^f_F(p)=\frac{-\FB{\cancel{p}+m_f}}{p^2-m_f^2+i\epsilon}
\end{eqnarray}
is the vacuum Feynman propagator for a spin-$\frac{1}{2}$ particle. 

In order to calculate the photon self energy at finite temperature, we use the standard techniques of RTF of TFT in which all the two point correlation functions such as propagator and self energy become $2\times2$ matrices in thermal indices~\cite{Bellac:2011kqa, Mallik:2016anp}. However, they can be put in diagonal forms in terms of analytic functions which we will denote by a bar. This also enables one to express the analytic function in terms any one component of the $2\times2$ matrix say the $11$ component. Denoting the $11$-component of thermal photon self energy matrix by $\Pi^\munu_{11}(q^0,\vec{q})$, we can write,
\begin{eqnarray}
\Pi^\munu_{11}(q^0,\vec{q}) = -\sum_{f}^{}e_f^2N_ci\int\frac{d^4k}{(2\pi)^4}
\Tr\TB{\gamma^\nu S_{11}^f(p)\gamma^\mu S_{11}^f(k)}
\label{eq.Pi.11.t}
\end{eqnarray}
where, $S^f_{11}(p)$ is the $11$-component of the real time Dirac propagator given by
\begin{eqnarray}
S^f_{11}(p) = S^f_F(p) - \eta(p\cdot u)\TB{S^f_F(p)-\gamma^0S^{f\dagger}_F(p)\gamma^0}~.
\label{eq.S11.1}
\end{eqnarray} 
In the above equation, $\eta(p\cdot u)=\Theta(p\cdot u)f(p\cdot u)+\Theta(-p\cdot u)f(-p\cdot u)$ with the quasi quark thermal distribution function $f$ defined in Eq.~(\ref{eq.distribution.function}). Here $u^\mu$ is the four velocity of the thermal medium which in local rest frame reduces to $u^\mu\equiv(1,\vec{0})$, $\beta$ is the inverse temperature and $\Theta(x)$ is the unit step function. It is to be noted that, the quark distribution function in the thermal propagator contains the effective fugacity parameter $z_q=z_q(T)$. 
%
As we already mentioned, $\Pi^\munu_{11}(q^0,\vec{q})$ is related to the analytic thermal self energy function of photon by the following relations~\cite{Bellac:2011kqa, Mallik:2016anp}
\begin{eqnarray}
\RE\overline{\Pi}^\munu(q^0,\vec{q}) &=& \RE\Pi^\munu_{11}(q^0,\vec{q}) \label{eq.analytic.relation.re} \\
\IM\overline{\Pi}^\munu(q^0,\vec{q}) &=& \sign{q^0}\tanh\FB{\frac{\beta q^0}{2}}\IM\Pi^\munu_{11}(q^0,\vec{q})
\label{eq.analytic.relation.im}
\end{eqnarray}
where $\sign{x}=\Theta(x)-\Theta(-x)$ is the signum function. We now rewrite Eq.(\ref{eq.S11.1}) as 
\begin{eqnarray}
S^f_{11}(p) = \FB{\cancel{p}+m_f}\TB{\FB{\frac{-1}{p^2-m_f^2+i\epsilon}}-\eta(p\cdot u)2\pi i\delta\FB{p^2-m_f^2}}~,
\label{eq.S11.2}
\end{eqnarray} 
and substitute into Eq.~(\ref{eq.Pi.11.t}). Performing the $dk^0$ integration and using Eqs.~(\ref{eq.analytic.relation.re}) and (\ref{eq.analytic.relation.im}) we obtain the real and imaginary part of the photon self energy function as
\begin{eqnarray}
\RE\overline{\Pi}^\munu(q^0,\vec{q}) &=& \RE\Pi_\text{vac}^\munu(q) + \sum_{f}^{}e_f^2N_c\bigintsss\frac{d^3k}{(2\pi)^3}\mathcal{P}
\TB{\frac{f(\omega_k^f)}{2\omega_k^f}\SB{\frac{N_f^\munu(k^0=-\omega_k^f)}{(q^0-\omega_k^f)^2-(\omega_p^f)^2}
+ \frac{N_f^\munu(k^0=\omega_k^f)}{(q^0+\omega_k^f)^2-(\omega_p^f)^2}} \right. \nn \\
&& \hspace{3.5cm} \left.
\frac{f(\omega_p^f)}{2\omega_p^f}\SB{\frac{N_f^\munu(k^0=-q^0-\omega_p^f)}{(q^0+\omega_p^f)^2-(\omega_k^f)^2}
+ \frac{N_f^\munu(k^0=-q^0+\omega_p^f)}{(q^0-\omega_p^f)^2-(\omega_k^f)^2}}} \label{eq.re.pibar.t}\\
\IM\overline{\Pi}^\munu(q^0,\vec{q}) &=& \sign{q^0}\tanh\FB{\frac{\beta q^0}{2}}\sum_{f}^{}e_f^2N_c
\pi\bigintsss\frac{d^3k}{(2\pi)^3}\frac{1}{4\omega_k^f\omega_p^f}
\TB{ \SB{1-f(\omega_k^f)-f(\omega_p^f)+2f(\omega_k^f)f(\omega_p^f)} \right. \nn \\
&& \hspace{-2.5cm}\left. \SB{N_f^\munu(k^0=-\omega_k^f)\delta(q^0-\omega_k^f-\omega_p^f)
	+N_f^\munu(k^0=\omega_k^f)\delta(q^0+\omega_k^f+\omega_p^f) } 
+ \SB{-f(\omega_k^f)-f(\omega_p^f)+2f(\omega_k^f)f(\omega_p^f)} \right. \nn \\
&& \left. \SB{N_f^\munu(k^0=-\omega_k^f)\delta(q^0-\omega_k^f+\omega_p^f)
	+N_f^\munu(k^0=\omega_k^f)\delta(q^0+\omega_k^f-\omega_p^f) }} \label{eq.im.pibar.t}
\end{eqnarray}
where $\mathcal{P}$ denotes the Cauchy principal value and 
\begin{eqnarray}
N^\munu_f(q,k) = \Tr\TB{\gamma^\nu\FB{\cancel{q}+\cancel{k}+m_f}\gamma^\mu\FB{\cancel{k}+m_f}} 
= 4\TB{(m_f^2-k^2-k\cdot q)g^\munu+2k^\mu k^\nu+(k^\mu q^\nu+k^\nu q^\mu)}~.
\label{eq.N.t}
\end{eqnarray}

First term on the R.H.S. of Eq.~\eqref{eq.re.pibar.t} is ultraviolet divergent. 
Performing dimensional regularization followed by $\overline{\text{MS}}$ scheme, we get~\cite{Peskin:1995ev}, 
\begin{eqnarray}
\RE\Pi_\text{vac}^\munu(q) = \FB{q^2g^\munu-q^\mu q^\nu}\sum_{f}^{}\frac{e_f^2N_c}{2\pi^2}\int_{0}^{1}dxx(x-1)\ln\MB{\frac{m_f^2-x(1-x)q^2}{\Lambda}} \label{eq.repi.vac}
\end{eqnarray}
where $\Lambda$ is a scale of dimension GeV$^2$. It is worth mentioning that, only the pure vacuum contribution contains the ultraviolet divergence whereas the temperature dependent part is free from any divergences~\cite{Bellac:2011kqa}. Analogous argument will also hold for the self energy in presence of external magnetic field where the divergent part will only come from the pure vacuum contribution whereas the magnetic field dependent terms will be finite. 
However, the imaginary part of the self energy is finite and scale independent. Thus the electromagnetic spectral function 
as well as DPR, obtained from the imaginary part will also be independent of scale and regularization scheme. On the other hand, 
the contribution of the scale dependent pure vacuum self energy to the Debye mass is zero which will be discussed in Sec.~\ref{sec.debye}.

\section{Photon Self Energy at Finite Temperature under external magnetic field}\label{sec.self.tb}

In presence of a constant external magnetic field $\vec{B}=B\hat{z}$, the $11$-component of quasi quark propagator becomes~\cite{DOlivo:2002omk} 
\begin{eqnarray}
S^f_{11}(p) = S^f_B(p) - \eta(p\cdot u)\TB{S^f_B(p)-\gamma^0S^{f\dagger}_B(p)\gamma^0}~.
\label{eq.S11.tb}
\end{eqnarray} 
where $S^f_B(p)$ is the momentum space Schwinger proper time propagator for a charged spin-$\frac{1}{2}$ particle given by~\cite{Schwinger:1951nm, Ayala:2004dx}
\begin{eqnarray}
S^f_B(p) &=& i\int_{0}^{\infty} ds \exp\TB{is\SB{p_\parallel^2+\frac{\tan(e_fBs)}{e_fBs}p_\perp^2-m_f^2}} 
\TB{\frac{}{}\FB{\cancel{p}_\parallel+m_f}\SB{1-\gamma^1\gamma^2\tan(e_fBs)}+\cancel{p}_\perp\sec^2(e_fBs)}~.
\label{eq.schwinger.1}
\end{eqnarray}
Since we are considering magnetic field along +ve z-direction, we decompose the metric tensor $g^\munu=\FB{g_\parallel^\munu+g_\perp^\munu}$ where $g_\parallel^\munu=diag(1,0,0,-1)$ and $g_\perp^\munu=diag(0,-1,-1,0)$ so that $p_\parallel^\mu=p_\nu g_\parallel^\munu$ and $p_\perp^\mu=p_\nu g_\perp^\munu$. It is worth mentioning that the corresponding coordinate space Schwinger propagator contains a translationally non-invariant phase factor. It can be shown that, for one-loop self energy graphs of neutral particles (like photon in our case) in which the loop particles are equally charged, the phase factor gets cancelled. In our case, we can work with the momentum space quasi quark propagator. The proper time integral in Eq.~(\ref{eq.schwinger.1}) can be performed, so that the propagator can be written as a sum over Landau levels as
\begin{eqnarray}
S^f_B(p) = -\sum_{n=0}^{\infty}\TB{ \frac{(-1)^ne^{-\alpha_p}D_n^f(p)}{p_\parallel^2-m_f^2-2n\MB{e_fB}+i\epsilon}}
\label{eq.sf.b}
\end{eqnarray}
where $\alpha_p = -\frac{p_\perp^2}{\MB{e_fB}}$ and 
\begin{eqnarray}
D_n^f(p) = \FB{\cancel{p}_\parallel+m_f}\TB{\SB{\frac{}{}1+\sign{e_f}i\gamma^1\gamma^2}L_n(2\alpha_p)  
-\SB{\frac{}{}1-\sign{e_f}i\gamma^1\gamma^2}L_{n-1}(2\alpha_p)} - 4\cancel{p}_\perp L^1_{n-1}(2\alpha_p)
\label{eq.Dn}
\end{eqnarray}
in which $L^a_n(z)$ are the generalized Laguerre polynomial with $L^a_{-1}(z)=0$. We now rewrite Eq.~(\ref{eq.S11.tb}) using Eq.~(\ref{eq.sf.b}) as
\begin{eqnarray}
S^f_{11}(p) = \sum_{n=0}^{\infty}(-1)^ne^{-\alpha_p}D_n^f(p)
\TB{\FB{\frac{-1}{p_\parallel^2-m_f^2-2n\MB{e_fB}+i\epsilon}}
-\eta(p\cdot u)2\pi i\delta\FB{p_\parallel^2-m_f^2-2n\MB{e_fB}}}~,
\end{eqnarray}
and substitute into Eq.~(\ref{eq.Pi.11.t}). After performing the $dk^0$ integral and using Eqs.~(\ref{eq.analytic.relation.re}) and (\ref{eq.analytic.relation.im}) we obtain the real and imaginary part of the photon self energy function at finite temperature under external magnetic field as,
\begin{eqnarray}
\RE\overline{\Pi}^\munu(q^0,\vec{q}) &=& \sum_{f}^{}e_f^2N_c
\sum_{l=0}^{\infty} \sum_{n=0}^{\infty}\bigintsss\frac{d^3k}{(2\pi)^3}\mathcal{P}
\TB{\frac{f(\omega_{k,l}^f)}{2\omega_{k,l}^f}\SB{\frac{N_{f,nl}^\munu(k^0=-\omega_{k,l}^f)}{(q^0-\omega_{k,l}^f)^2-(\omega_{p,n}^f)^2}
		+ \frac{N_{f,nl}^\munu(k^0=\omega_{k,l}^f)}{(q^0+\omega_{k,l}^f)^2-(\omega_{p,n}^f)^2}} \right. \nn \\
	&& \hspace{0.0cm} \left.
	\frac{f(\omega_{p,n}^f)}{2\omega_{p,n}^f}\SB{\frac{N_{f,nl}^\munu(k^0=-q^0-\omega_{p,n}^f)}{(q^0+\omega_{p,n}^f)^2-(\omega_{k,l}^f)^2}
		+ \frac{N_{f,nl}^\munu(k^0=-q^0+\omega_{p,n}^f)}{(q^0-\omega_{p,n}^f)^2-(\omega_{k,l}^f)^2}}} 
	+ \RE\Pi_\text{vac}^\munu(q) +\RE\Pi_\text{B}^\munu(q,B)	
	 \label{eq.re.pibar.tb}\\
\IM\overline{\Pi}^\munu(q^0,\vec{q}) &=& \sign{q^0}\tanh\FB{\frac{\beta q^0}{2}}\sum_{f}^{}e_f^2N_c
\sum_{l=0}^{\infty} \sum_{n=0}^{\infty}\pi\bigintsss\frac{d^3k}{(2\pi)^3}\frac{1}{4\omega_{k,l}^f\omega_{p,n}^f}
\TB{ \SB{1-f(\omega_{k,l}^f)-f(\omega_{p,n}^f)+2f(\omega_{k,l}^f)f(\omega_{p,n}^f)} \right. \nn \\
	&& \hspace{-2.5cm}\left. \SB{N_{f,nl}^\munu(k^0=-\omega_{k,l}^f)\delta(q^0-\omega_{k,l}^f-\omega_{p,n}^f)
		+N_{f,nl}^\munu(k^0=\omega_{k,l}^f)\delta(q^0+\omega_{k,l}^f+\omega_{p,n}^f) } 
	+ \SB{-f(\omega_{k,l}^f)-f(\omega_{p,n}^f)+2f(\omega_{k,l}^f)f(\omega_{p,n}^f)} 
	\right. \nn \\	&& \left. 
	\SB{N_{f,nl}^\munu(k^0=-\omega_{k,l}^f)\delta(q^0-\omega_{k,l}^f+\omega_{p,n}^f)
		+N_{f,nl}^\munu(k^0=\omega_{k,l}^f)\delta(q^0+\omega_{k,l}^f-\omega_{p,n}^f) }} \label{eq.im.pibar.tb}
\end{eqnarray}
where 
\begin{eqnarray}
\omega_{k,l}^f = \sqrt{k_z^2+m_{f,l}^2} ~~~ \text{with} ~~ m_{f,l}=\sqrt{m_f^2+2n\MB{e_fB}} \label{eq.mfl}~~ \text{and}\\
N_{f,nl}^\munu(q,k) = (-1)^{n+l}e^{-\alpha_k-\alpha_p}\Tr\TB{\gamma^\nu D_n^f(p)\gamma^\mu D_l^f(k)}~.
\label{eq.Nnl}
\end{eqnarray}

It is worth mentioning that, the last term in Eq.~(\ref{eq.re.pibar.tb}) i.e. $\RE\Pi_\text{B}^\munu(q,B)$ is the magnetic field dependent vacuum contribution to the real part of the self energy which is temperature independent. This term will contribute to the dispersion relations of photon. Similar study incorporating this magnetic field dependent vacuum contribution term, for the dispersion relations of $\pi$ and $\rho$ mesons, can be found in Refs.~\cite{Mukherjee:2017dls, Ghosh:2017rjo, Ghosh:2016evc}. However, in this work, we have not given the explicit calculation of this term since it does not contribute to the Debye mass which we will discuss in the next section.


\section{Debye Screening Mass}\label{sec.debye}
The Debye screening mass $m_D$ for electromagnetic screening can be obtained by taking static limit of the $00$-component of photon polarization tensor,
\begin{eqnarray}
m_D^2 = - \RE\overline{\Pi}^{00}\FB{q^0=0,\vec{q}\rightarrow\vec{0}}~.
\end{eqnarray}
Let us first calculate $m_D$ for \textit{zero magnetic field} case. From Eq.~(\ref{eq.re.pibar.t}) we get,
\begin{eqnarray}
\lim\limits_{q^0=0,\vec{q}\rightarrow\vec{0}}\RE\overline{\Pi}^{00}(q^0,\vec{q}) 
= -\FB{\frac{4N_c}{T}}\sum_{f}^{}e_f^2\int\frac{d^3k}{(2\pi)^3}f(\omega_k^f)\TB{1-f(\omega_k^f)} = -m_D^2~.
\label{eq.md}
\end{eqnarray}
Here, the contribution from the scale dependent pure vacuum self energy given in Eq.~\eqref{eq.repi.vac} vanishes 
as $\lim\limits_{q^0=0,\vec{q}\rightarrow\vec{0}} \RE\Pi_\text{vac}^\munu(q) = 0$ so that the Debye mass becomes independent of 
scale and regularization scheme.
The above integral can be analytically evaluated for massless quarks ($m_f=0$) and we get,
\begin{eqnarray}
m_D^2= -\FB{\frac{4N_cT^2}{\pi^2}}\text{Li}_2(-z_q)\sum_{f}^{}e_f^2 ~~~~ \text{for massless quarks}
\label{eq.md.t}
\end{eqnarray} 
where $\text{Li}_2(z)$ is the dilogarithm function. The above equation reduces to the well known expression of Debye mass 
\begin{eqnarray}
m_D^2= N_c\sum_{f}^{}\FB{\frac{e_f^2T^2}{3}}  ~~~ \text{for the ideal case~} z_q=1.
\label{eq.md.ideal}
\end{eqnarray}

Similar result can be obtained from semi classical transport theory~\cite{Jankowski:2015eoa}
\begin{eqnarray}
-m_D^2 &=& g\sum_{f}^{}e_f^2\int\frac{d^3k}{(2\pi)^3}\frac{\partial}{\partial\omega_k^f}f(\omega_k^f) \nn \\
&=& -g\sum_{f}^{}e_f^2\frac{1}{T}\int\frac{d^3k}{(2\pi)^3}f(\omega_k^f)\TB{1-f(\omega_k^f)} \label{eq.md.sctt}
\end{eqnarray}
where $g=2\times 2\times N_c$ is the degeneracy factor for quark-antiquark, spin and colour. It can be easily checked that the above equation boils down to Eq.~(\ref{eq.md}).

Let us now turn on the \textit{external magnetic field}. In this case we will take static limit to Eq.~(\ref{eq.re.pibar.tb}) details of which are provided in Appendix~\ref{app.repi00}. The Debye mass in presence of external magnetic field comes out to be
\begin{eqnarray}
m_D^2 = \FB{\frac{N_cB}{\pi^2}}\sum_{f}^{}\MB{e_f}^3\sum_{n=0}^{\infty}(2-\delta_n^0)
\int_{0}^{\infty}\frac{dk_z}{T}f(\omega_{k,n}^f)\TB{1-f(\omega_{k,n}^f)}~.
\label{eq.md.b}
\end{eqnarray}

Analogous results can be found from the semi classical transport theory as well. We have the expression for Debye mass from Eq.~(\ref{eq.md.sctt}) as 
\begin{eqnarray}
m_D^2 = g\sum_{f}^{}e_f^2\frac{1}{T}\int\frac{d^3k}{(2\pi)^3}f(\omega_k^f)\TB{1-f(\omega_k^f)}~.
\label{eq.md.b.sctt}
\end{eqnarray}
In this case, due to Landau quantization of the transverse momentum $k_\perp^2=-2n\MB{e_fB}$ of quarks/antiquarks, the dispersion relation becomes 
\begin{eqnarray}
\omega_k^f \longrightarrow \omega_{k,n}^f = \sqrt{k_z^2+m_f^2+2n\MB{e_fB}}~.
\end{eqnarray}
This in turn modifies the phase space integration in Eq.~\ref{eq.md.b.sctt} as
\begin{eqnarray}
\int\frac{d^3k}{(2\pi)^3} \longrightarrow
\frac{\MB{e_fB}}{(2\pi)^3}\sum_{n=0}^{\infty}\int_{0}^{2\pi}d\phi\int_{-\infty}^{\infty}dk_z
\end{eqnarray}
so that we get from Eq.~(\ref{eq.md.b.sctt})
\begin{eqnarray}
m_D^2 = \FB{\frac{B}{2\pi^2}}\sum_{f}^{}\MB{e_f}^3\sum_{n=0}^{\infty}g_n
\int_{0}^{\infty}\frac{dk_z}{T}f(\omega_{k,n}^f)\TB{1-f(\omega_{k,n}^f)}
\end{eqnarray}
where the degeneracy factor $g_n$ is dependent on the Landau level index $n$. It is well known~\cite{Bhattacharya:2007vz} that the Lowest Landau Level (LLL) is spin non-degenerate which implies that $g_n=2\times(2-\delta_n^0)\times N_c$. Substituting $g_n$ in the above expression we recover Eq.~(\ref{eq.md.b}).

\section{Electromagnetic Spectral Function} \label{sec.rho}
The electromagnetic spectral function $\rho^{q\bar{q}}(q)$ due to quark loop is defined as 
\begin{eqnarray}
\rho^{q\bar{q}}(q) = \frac{1}{4\pi\alpha} g_\munu \IM \overline{\Pi}^\munu(q)~. \label{eq.rho.def}
\end{eqnarray}
Let us first calculate $\rho^{q\bar{q}}(q)$ for \textit{zero magnetic field case} which is obtained from Eq.~(\ref{eq.im.pibar.t}) after contracting with $g_\munu$. The four terms in Eq.~(\ref{eq.im.pibar.t}) with the four Dirac delta functions are termed as Unitary-I, Unitary-II, Landau-II and Landau-I cuts respectively according to their appearance in that equation. These terms are non-vanishing in certain kinematic domains as can be read from Appendix~\ref{app.kinematic.domain}. The Unitary-I term is non-vanishing for $\sqrt{\vec{q}^2+4m_f^2} \le q^0 < \infty$ whereas Unitary-II term is non-vanishing for $-\infty < q^0 \le -\sqrt{\vec{q}^2+4m_f^2}$. Both the Landau terms have their corresponding kinematic region as $|q^0| \le |\vec{q}|$. These different cuts correspond to different physical processes like decay and scattering. For example Unitary-I cut correspond to the decay of a photon with energy $q^0>\sqrt{\vec{q}^2+4m_f^2}$ into a quark-antiquark pair (which is the threshold energy for a pair creation) and also the time reversed process that a quark-antiquark annihilate to make a photon. Similarly the Landau cuts correspond to the absorption of a photon  with energy $|q^0| \le |\vec{q}|$ due to scattering with a quark producing another quark in the final state and also the time reversed process that a quark emits a photon in the medium. If we restrict ourselves to the physical time like region defined in terms of $q^0>0$ and $q^2>0$, then only the Unitary-I cut contributes. Therefore, in the physical region, processes like photon decay and formation occurs via the Unitary-I cut where as the scattering and emission processes do not occur.

We then evaluate the $d(\cos\theta)$ integrals in Eq.~(\ref{eq.im.pibar.t}) by using the Dirac delta functions~\cite{Mallik:2016anp} in order to simplify the expression of the spectral function and impose the kinematic restrictions discussed above to get, 
\begin{eqnarray}
\rho^{q\bar{q}}(q^0,\vec{q}) &=& \frac{1}{4\pi\alpha}\sign{q^0}\tanh\FB{\frac{\beta q^0}{2}}\sum_{f}^{}e_f^2N_c
\frac{1}{16\pi|\vec{q}|}\TB{\int_{\omega^f_-}^{\omega^f_+}d\omega_k^f U_1^f(\cos\theta=\cos\theta_0)
\Theta\FB{q^0-\sqrt{\vec{q}^2+4m_f^2}} \right. \nn \\ && \left.
+\int_{-\omega^f_+}^{-\omega^f_-}d\omega_k^f U_2^f(\cos\theta=\cos\theta_0')
\Theta\FB{-q^0-\sqrt{\vec{q}^2+4m_f^2}} +
\int_{-\omega^f_+}^{\infty}d\omega_k^f L_1^f(\cos\theta=\cos\theta_0')\Theta\FB{-|q^0|+|\vec{q}|}
\right. \nn \\ && \left.
+ \int_{\omega^f_-}^{\infty}d\omega_k^f L_2^f(\cos\theta=\cos\theta_0)\Theta\FB{-|q^0|+|\vec{q}|}}
\label{eq.rho.t}
\end{eqnarray}
where, 
\begin{eqnarray}
\omega^f_\pm &=& \frac{1}{2q^2}\TB{q^0q^2\pm|\vec{q}|\lambda^{1/2}\FB{q^2,m_f^2,m_f^2}} \\
U_1 &=& \SB{1-f(\omega_k^f)-f(\omega_p^f)+2f(\omega_k^f)f(\omega_p^f)}g_\munu N_f^\munu (k^0=-\omega_k^f) \\
U_2 &=& \SB{1-f(\omega_k^f)-f(\omega_p^f)+2f(\omega_k^f)f(\omega_p^f)}g_\munu N_f^\munu (k^0=\omega_k^f) \\
L_1 &=& \SB{-f(\omega_k^f)-f(\omega_p^f)+2f(\omega_k^f)f(\omega_p^f)}g_\munu N_f^\munu (k^0=\omega_k^f) \\
L_2 &=& \SB{-f(\omega_k^f)-f(\omega_p^f)+2f(\omega_k^f)f(\omega_p^f)}g_\munu N_f^\munu (k^0=-\omega_k^f) \\
\cos\theta_0 &=& \FB{\frac{-2q^0\omega_k^f+q^2}{2|\vec{q}||\vec{k}|}} \\
\cos\theta_0' &=& \FB{\frac{2q^0\omega_k^f+q^2}{2|\vec{q}||\vec{k}|}}~.
\end{eqnarray}
with $\lambda(x,y,z)=x^2+y^2+z^2-2xy-2yz-2zx$ being the K\"all\'en function. It is easy to check from Eq.~(\ref{eq.N.t}) that $g_\munu N_f^\munu = 8\FB{2m_f^2-k^2-k\cdot q}$.

Let us now turn on the \textit{external magnetic field}. In this case, the calculation spectral function is simplified if we consider the transverse momentum $q_\perp$ of the photon to be zero. This has been provided in Appendix~\ref{app.simplification.rho} and we get from Eq.~(\ref{eq.rho.tb2})
\begin{eqnarray}
\rho^{q\bar{q}}(q^0,q_z) &=& \frac{1}{4\pi\alpha}\sign{q^0}\tanh\FB{\frac{\beta q^0}{2}}\sum_{f}^{}e_f^2N_c
\sum_{n=0}^{\infty} ~\sum_{l=(n-1)}^{(n+1)}\pi\bigintsss_{-\infty}^{\infty}
\frac{dk_z}{2\pi}\frac{1}{4\omega_{k,l}^f\omega_{p,n}^f}
\TB{\tilde{U}^f_{1,nl}\delta\FB{q^0-\omega_{k,l}^f-\omega_{p,n}^f} \right. \nn \\ && \left.
	+\tilde{U}^f_{2,nl}\delta\FB{q^0+\omega_{k,l}^f+\omega_{p,n}^f}
	+\tilde{L}^f_{1,nl}\delta\FB{q^0+\omega_{k,l}^f-\omega_{p,n}^f}
	+ \tilde{L}^f_{2,nl}\delta\FB{q^0-\omega_{k,l}^f+\omega_{p,n}^f}}
\label{eq.rho.tb2.dual}~.
\end{eqnarray}

Let us now discuss the analytic structure of the spectral function in presence of external magnetic field. As can be obtained from Appendix~\ref{app.kinematic.domain}, in this case, the kinematic regions for Unitary-I and Unitary-II cuts are respectively $\sqrt{q_z^2+4m_f^2} \le q^0 < \infty$ and $-\infty < q^0 \le -\sqrt{q_z^2+4m_f^2}$ whereas the corresponding region for both the Landau cuts is $|q^0|\le \sqrt{q_z^2+\FB{m_f-\sqrt{m_f^2+2|e_fB|}}^2}$. Hence, if we restrict ourselves to the physical time like regions in terms of $q^0>0$ and $q^2>0$, then along with Unitary-I cut, both the Landau cuts contribute. The appearance of these Landau cuts is a purely magnetic field effect as can be noticed that, at $B\rightarrow 0$, the Landau cuts disappear from the physical time like region. Physically this means that, in addition to the decay/formation processes, scattering/emission can also happen in presence of external magnetic field. A photon can scatter with a quark from lower Landau level to get absorbed producing another quark on a higher Landau level and the time reversed process like a quark from a higher Landau level can emit a photon and goes down to a lower Landau level. As mentioned earlier, this type of process can not happen in absence of external magnetic field. We will see in the next section, that the appearance of the Landau cuts will enhance the low invariant mass dilepton production rate.

Let us now simplify Eq.~(\ref{eq.rho.tb2.dual}) by evaluating the $dk_z$ integral using the Dirac delta functions and impose the kinematic restrictions as discussed above. The result is
\begin{eqnarray}
\rho^{q\bar{q}}(q^0,q_z) &=& \frac{1}{4\pi\alpha}\sign{q^0}\tanh\FB{\frac{\beta q^0}{2}}\sum_{f}^{}e_f^2N_c
\sum_{n=0}^{\infty} ~\sum_{l=(n-1)}^{(n+1)}\frac{1}{4\lambda^{1/2}(q_\parallel^2,m_{f,l}^2,m_{f,n}^2)} \nn \\ && \sum_{\tilde{k}_z\in\tilde{k}_z^\pm}^{}
\TB{\frac{}{}\tilde{U}^f_{1,nl}(k_z=\tilde{k}_z)	\Theta\FB{q^0-\sqrt{q_z^2+(m_{f,l}+m_{f,n})^2}}
+ \tilde{U}^f_{2,nl}(k_z=\tilde{k}_z) \Theta\FB{-q^0-\sqrt{q_z^2+(m_{f,l}+m_{f,n})^2}}
\right. \nn \\ && \hspace{2cm}\left.
+ \tilde{L}^f_{1,nl}(k_z=\tilde{k}_z) \Theta\FB{q^0-\min\FB{q_z,E'_\pm}}\Theta\FB{-q^0+\max\FB{q_z,E'_\pm}}
\right. \nn \\ && \hspace{2cm}\left.	 
+ \tilde{L}^f_{2,nl}(k_z=\tilde{k}_z) \Theta\FB{-q^0-\min\FB{q_z,E'_\pm}}\Theta\FB{q^0+\max\FB{q_z,E'_\pm}\frac{}{}}
}
\label{eq.rho.tb3}
\end{eqnarray}
where, $\tilde{k}_z^\pm = \frac{1}{2q_\parallel^2}\TB{-yq_z\pm|q^0|\lambda^{1/2}\FB{q_\parallel^2,m_{f,l}^2,m_{f,n}^2}}$, 
$y=(q_\parallel^2+m_{f,l}^2-m_{f,n}^2)$, $\tilde{\omega}_{k,l}^f = \sqrt{\tilde{k}_z^2+m_{f,l}^2}$, ~~~
$E'_\pm = \frac{m_{f,l}-m_{f,n}}{\MB{m_{f,l}\pm m_{f,n}}}\sqrt{q_z^2+(m_{f,l}\pm m_{f,n})^2}$ 
and $m_{f,l}$ is defined in Eq.~\eqref{eq.mfl}.

\section{Dilepton Production Rate at zero external magnetic field}\label{sec.dlr.0}

In order to calculate the DPR from hot QCD medium, we follow the standard procedure given in Ref.~\cite{Mallik:2016anp} and consider an initial state $\ket{i}=\ket{I(p_I)}$ of quark/antiquark with momentum $p_I$, which goes to a final state $\ket{f}=\ket{F(p_F)l^+(p_+,s_+)l^-(p_-,s_-)}$ containing the quark/antiquark with momentum $p_F$ plus dilepton with momenta $p_+,p_-$ and spin $s_+,s_-$. The probability amplitude for transition $\ket{i}\longrightarrow\ket{f}$ is $\expectationvalue{f}{\hat{S}}{i}$ where $\hat{S}$ is the scattering matrix operator 
\begin{eqnarray}
\hat{S} = \mathcal{T}\TB{\exp\SB{i\int\scrL_\text{int}(x)d^4x}}
\end{eqnarray}
with
\begin{eqnarray}
\scrL_\text{int}(x) = \TB{\frac{}{}j^\mu(x)+J^\mu(x)}A_\mu(x)~.
\end{eqnarray}
Here $\mathcal{T}$ is the time ordering symbol, $j^\mu(x)$ is the lepton current, $J^\mu(x)$ is the quark/antiquark current and $A_\mu(x)$ is the photon field. The lepton and quark/antiquark currents are given by
\begin{eqnarray}
j^\mu(x) &=& -e\bar{\psi}(x)\gamma^\mu\psi(x) \label{eq.lepton.current.0}\\
J^\mu(x) &=& \sum_{f}^{}e_f\bar{q}_f(x)\gamma^\mu q_f(x) \label{eq.quark.current.0}
\end{eqnarray}
where, $\psi(x)$ and $q_f(x)$ are respectively the lepton and quark fields. Expanding $\hat{S}$ up to second order and after some simplifications, the non-trivial contribution to the $S$-matrix element comes from 
\begin{eqnarray}
\expectationvalue{f}{\hat{S}}{i} = -\int\int d^4x_1 d^4x_2 \expectationvalue{F}{J_\mu(x_1)}{I}\expectationvalue{l^+l^-}{j_\nu(x_2)}{0}\Delta_F^\munu(x_1-x_2)
\label{eq.smatrix.1}
\end{eqnarray}
where, $\Delta_F^\munu(x_1-x_2)=\expectationvalue{0}{\mathcal{T}A^\mu(x_1)A^\nu(x_2)}{0}$ is the coordinate space vacuum Feynman photon propagator which can be Fourier transformed as 
\begin{eqnarray}
\Delta_F^\munu(x_1-x_2) = \int\frac{d^4q}{(2\pi)^4}e^{-iq\cdot(x_1-x_2)}(-i\Delta_F^\munu(q)) 
\label{eq.phonon.propagator.x}
\end{eqnarray} 
in which 
\begin{eqnarray}
\Delta_F^\munu(q) = -g^\munu\FB{\frac{-1}{q^2+i\epsilon}}
\label{eq.phonon.propagator.p}
\end{eqnarray}
is the momentum space vacuum Feynman photon propagator.
\begin{figure}[h]
	\begin{center}
		\includegraphics[angle=0, scale=0.45]{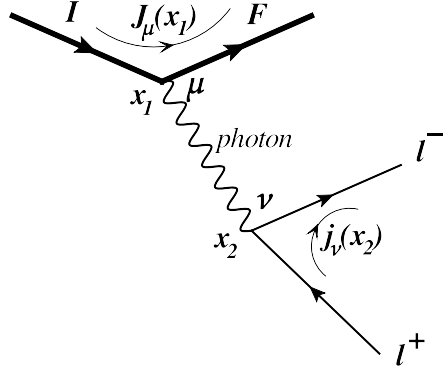}
	\end{center}
	\caption{Diagrammatic representation of the dilepton production amplitude from Eq.~(\ref{eq.smatrix.1}).}
	\label{fig.dilepton.amplitude}
\end{figure}
Fig.~\ref{fig.dilepton.amplitude} shows the diagrammatic representation of the dilepton production amplitude as given in  Eq.~(\ref{eq.smatrix.1}). In Eq.~(\ref{eq.smatrix.1}), a translation of the leptonic current operator $j_\nu(x_2)$ to the origin leads to the following simplification of its matrix element
\begin{eqnarray}
\expectationvalue{l^+l^-}{j_\nu(x_2)}{0}
=\expectationvalue{l^+l^-}{e^{i\hat{P}\cdot x_2}j_\nu(0)e^{-i\hat{P}\cdot x_2}}{0}
= e^{i(p_++p_-)\cdot x_2}\expectationvalue{l^+l^-}{j_\nu(0)}{0}~.
\label{eq.lepton.current}
\end{eqnarray}
We now substitute Eq.~(\ref{eq.phonon.propagator.x}) and (\ref{eq.lepton.current}) into Eq.~(\ref{eq.smatrix.1}) and perform the $d^4q$ integral which gives a Dirac delta function $\delta^4(q+p_++p_-)$. This delta function is in turn used to perform the $d^4x_2$ integral and we are left with
\begin{eqnarray}
\expectationvalue{f}{\hat{S}}{i} = i\int d^4x_1 e^{i(p_++p_-)\cdot x_1}\expectationvalue{F}{J_\mu(x_1)}{I}
\expectationvalue{l^+l^-}{j_\nu(0)}{0}\Delta_F^\munu(-p_+-p_-)~.
\end{eqnarray} 
Taking absolute square of the above equation, we get
\begin{eqnarray}
\MB{\expectationvalue{f}{\hat{S}}{i}}^2 &=& \int\int d^4x_1d^4xe^{i(p_++p_-)\cdot(x_1-x)}
\expectationvalue{F}{J_\mu(x_1)}{I}\expectationvalue{I}{J^\dagger_\alpha(x)}{F} \nn \\
&& \expectationvalue{l^+l^-}{j_\nu(0)}{0} \expectationvalue{0}{j^\dagger_\beta(0)}{l^+l^-}
\Delta_F^\munu(-p_+-p_-)\Delta_F^\alphabeta(-p_+-p_-)~.
\label{eq.smatrix.2}
\end{eqnarray}
Translating the quark/antiquark current operator $J^\dagger_\alpha(x)$ to the origin, we can write its matrix element as 
\begin{eqnarray}
\expectationvalue{I}{J^\dagger_\alpha(x)}{F}
= e^{i(p_I-p_F)\cdot x}\expectationvalue{I}{J^\dagger_\alpha(0)}{F}~.
\end{eqnarray}
We now substitute the above matrix element as well as the momentum space photon propagator from Eq.~(\ref{eq.phonon.propagator.p}) into Eq.~(\ref{eq.smatrix.2}) and impose the momentum conservation 
$p_I=p_F+p_++p_-$ to get,
\begin{eqnarray}
\MB{\expectationvalue{f}{\hat{S}}{i}}^2 = \int\int d^4x_1d^4x\frac{e^{i(p_++p_-)\cdot x_1}}{(p_++p_-)^4}
\expectationvalue{F}{J_\mu(x_1)}{I}\expectationvalue{I}{J^\dagger_\nu(0)}{F} 
\expectationvalue{l^+l^-}{j^\mu(0)}{0} \expectationvalue{0}{j^{\dagger\nu}(0)}{l^+l^-}
\label{eq.smatrix.3}
\end{eqnarray}
The dilepton multiplicity ($N$) from the thermal QGP medium is given by~\cite{Mallik:2016anp}
\begin{eqnarray}
N = \sum_{s_+,s_-}^{}\int\frac{d^3p_+}{(2\pi)^32E_+}\int\frac{d^3p_-}{(2\pi)^32E_-}\frac{1}{\mathcal{Z}}
\sum_{I,F}^{}e^{-\beta E_I}\MB{\expectationvalue{f}{\hat{S}}{i}}^2
\end{eqnarray}
where, $E_\pm=\sqrt{m^2+\vec{p}_\pm^{~2}}$, $m$ is the lepton mass and $\mathcal{Z}$ is the partition function of the system. We now substitute Eq.~(\ref{eq.smatrix.3}) into the 
above equation and get,
\begin{eqnarray}
N &=& \sum_{s_+,s_-}^{}\int\frac{d^3p_+}{(2\pi)^32E_+}\int\frac{d^3p_-}{(2\pi)^32E_-}
\expectationvalue{l^+l^-}{j^\mu(0)}{0} \expectationvalue{0}{j^{\dagger\nu}(0)}{l^+l^-}
\int\int d^4xd^4x_1\frac{e^{i(p_++p_-)\cdot x_1}}{(p_++p_-)^4} \times \nn \\ && \hspace{2cm}
\TB{ \frac{1}{\mathcal{Z}}\sum_{I,F}^{}e^{-\beta E_I}
\expectationvalue{F}{J_\mu(x_1)}{I}\expectationvalue{I}{J^\dagger_\nu(0)}{F}}~.
\label{eq.dlp.N}
\end{eqnarray}
The quantity within the square bracket in the above equation can be simplified using the momentum conservation $p_I=p_F+p_++p_-$ and the completeness relation of the initial states $\sum_{I}^{}\outerproduct{I}{I}=\mathds{1}$ so that,
\begin{eqnarray}
\frac{1}{\mathcal{Z}}\sum_{I,F}^{}e^{-\beta E_I}
\expectationvalue{F}{J_\mu(x_1)}{I}\expectationvalue{I}{J^\dagger_\nu(0)}{F}
= e^{-\beta(E_++E_-)}\ensembleaverage{J_\mu(x_1)J_\nu^\dagger(0)}
\end{eqnarray}
where, $\ensembleaverage{}$ denotes the ensemble average. Substituting the above equation into Eq.~(\ref{eq.dlp.N}) and inserting the identity $1=\int d^4q\delta^4(q-p_+-p_-)$ in the R.H.S. of Eq.~(\ref{eq.dlp.N}) 
we get
\begin{eqnarray}
N = \int\int d^4xd^4q \frac{e^{-\beta q^0}}{(2\pi)^4q^4} W^+_\munu(q) L^{+\munu}(q)
\label{eq.dlp.N.1}
\end{eqnarray}
where, 
\begin{eqnarray}
W^+_\munu(q) &=& \int d^4x e^{iq\cdot x} \ensembleaverage{J_\mu(x)J_\nu^\dagger(0)} \label{eq.W+.1}\\
L^+_\munu(q) &=& (2\pi)^4\sum_{s_+,s_-}^{}\int\frac{d^3p_+}{(2\pi)^32E_+}\int\frac{d^3p_-}{(2\pi)^32E_-}
\delta^4(q-p_+-p_-)\expectationvalue{l^+l^-}{j_\mu(0)}{0} \expectationvalue{0}{j^\dagger_\nu(0)}{l^+l^-}~.
\label{eq.L+.1}
\end{eqnarray}
Thus, the DPR i.e. the dilepton production per unit phase space volume $d^4xd^4q$ is given by (from Eq.~(\ref{eq.dlp.N.1}))
\begin{eqnarray}
\text{DPR} = \FB{\frac{dN}{d^4qd^4x}} = \frac{e^{-\beta q^0}}{(2\pi)^4q^4} W^+_\munu(q) L^{+\munu}(q)~.
\label{eq.dpr.1}
\end{eqnarray}

In order to calculate $W^+_\munu(q)$, we use the RTF of TFT in which it is related to the imaginary part of the $11$-component of the Fourier transform of time order current-current correlator~\cite{Bellac:2011kqa,Mallik:2016anp},
\begin{eqnarray}
W^+_\munu(q) = \FB{\frac{1}{e^{-\beta q^0}+1}}2~\IM~W^{11}_\munu(q)
\label{eq.W+.ImW11}
\end{eqnarray}
where, 
\begin{eqnarray}
W^{11}_\munu(q) = i\int d^4x e^{iq\cdot x}\ensembleaverage{\mathcal{T}J_\mu(x)J_\nu^\dagger(0)}~.
\label{eq.w11}
\end{eqnarray}
The calculation of $W^{11}_\munu(q)$ is provided in the Appendix~\ref{appendix.w11} and the final result can be read of from Eq.~(\ref{eq.w11.3}) which comes out to be exactly the (-1) times the $11$-component of the real time thermal photon self energy given in Eq.~(\ref{eq.Pi.11.t}). Thus the DPR becomes 
\begin{eqnarray}
\text{DPR} = \FB{\frac{dN}{d^4qd^4x}} = \frac{e^{-\beta q^0}}{(2\pi)^4q^4} \FB{\frac{-2}{e^{-\beta q^0}+1}}
\IM\Pi^{11}_\numu(q) L^{+\munu}(q)~.
\label{eq.dpr.2}
\end{eqnarray}

The lepton tensor $L^+_\munu(q)$ on Eq.~(\ref{eq.L+.1}) can be expressed in a more convenient form which is analogous to $W^+_\munu(q)$ in Eq.~(\ref{eq.W+.1}). For that, we write
\begin{eqnarray}
(2\pi)^4\delta^4(q-p_+-p_-)\expectationvalue{0}{j^\dagger_\nu(0)}{l^+l^-} = 
\int d^4x e^{ix\cdot(q-p_+-p_-)}\expectationvalue{0}{j^\dagger_\nu(0)}{l^+l^-} 
= \int d^4x e^{iq\cdot x} \expectationvalue{0}{j^\dagger_\nu(x)}{l^+l^-}
\end{eqnarray}
where the lepton current operator $j^\dagger_\nu(0)$ has been translated from origin to $x$. Substituting the above equation into Eq.~(\ref{eq.L+.1}) we get,
\begin{eqnarray}
L^+_\munu(q) = \int d^4xe^{iq\cdot x} \sum_{s_+,s_-}^{}\int\frac{d^3p_+}{(2\pi)^32E_+}\int\frac{d^3p_-}{(2\pi)^32E_-}
\expectationvalue{0}{j^\dagger_\nu(x)}{l^+l^-}\expectationvalue{l^+l^-}{j_\mu(0)}{0} ~. 
\label{eq.L+.2}
\end{eqnarray}
We now use the completeness relation for the two-particle leptonic state
\begin{eqnarray}
\mathds{1} = \sum_{s_+,s_-}^{}\int\frac{d^3p_+}{(2\pi)^32E_+}\int\frac{d^3p_-}{(2\pi)^32E_-}
\outerproduct{l^+l^-}{l^+l^-} 
\end{eqnarray}
so that, Eq.~(\ref{eq.L+.2}) becomes
\begin{eqnarray}
L^+_\munu(q) = \int d^4xe^{iq\cdot x} \expectationvalue{0}{j^\dagger_\nu(x)j_\mu(0)}{0} ~. 
\label{eq.L+.3}
\end{eqnarray}
which is analogous to the expression of $W^+_\munu$ in Eq.~(\ref{eq.W+.1}) where we have ensemble average instead of the vacuum expectation value. Thus, similar to Eq.~(\ref{eq.W+.ImW11}), $L^+_\munu(q)$ in the above equation can be related to the imaginary part of the Fourier transform of the time order leptonic current-current correlator~\cite{Mallik:2016anp,Bellac:2011kqa} as
\begin{eqnarray}
L^+_\munu(q) = 2~\IM~L_\munu(q)
\label{eq.L+.ImL}
\end{eqnarray}
where, 
\begin{eqnarray}
L_\munu(q) = i\int d^4x e^{iq\cdot x}\expectationvalue{0}{\mathcal{T}j^\dagger_\nu(x)j_\mu(0)}{0}~.
\label{eq.L}
\end{eqnarray}
The calculation of $L_\munu$ is provided in the Appendix~\ref{appendix.L} and it comes out to be exactly the (-1) times the photon vacuum self energy $\pi_\munu(q)$ due to $l^+l^-$ loop given by
\begin{eqnarray}
-L_\munu(q) = \pi^\munu(q) = -ie^2\int\frac{d^4k}{(2\pi)^4}\Tr\TB{\gamma^\nu S_l(q+k)\gamma^\mu S_l(k)}
\label{eq.pi.1}
\end{eqnarray} 
where $S_l(p)=\frac{-(\cancel{p}+m)}{p^2-m^2+i\epsilon}$ is the vacuum lepton Feynman propagator. It is worth mentioning that both the form of $L^+_\munu(q)$ in Eqs.~(\ref{eq.L+.2}) and (\ref{eq.L+.ImL}) boil down to the same analytical expression which is shown in Appendix~\ref{appendix.2Ls}. The calculation of $L^+_\munu(q)$ from Eq.~(\ref{eq.L+.2}) requires the knowledge of spin sums over leptonic spinors whereas its calculation from (\ref{eq.L+.ImL}) does not require the same. We will see that, the particular form of $L^+_\munu(q)$ in Eq.~(\ref{eq.L+.ImL}) together with Eq.~(\ref{eq.pi.1}) will be of more convenience while introducing the external magnetic field. For that case, the charged lepton propagators $S_l(p)$ in the above equation will be replaced by the Schwinger proper-time one and thus a knowledge of spin sum in presence of external magnetic field will not be required.

We now substitute $L^+_\munu(q)$ from Eqs.(\ref{eq.L+.ImL}) and (\ref{eq.pi.1}) into Eq.~(\ref{eq.dpr.2}) followed by using Eq.~(\ref{eq.analytic.relation.im}) to obtain, 
\begin{eqnarray}
\text{DPR} = \FB{\frac{dN}{d^4qd^4x}} = \frac{4}{(2\pi)^4q^4} \FB{\frac{1}{e^{\beta q^0}-1}}
\IM\overline{\Pi}_\numu(q)~\IM\pi^\munu(q)~.
\label{eq.dpr.3}
\end{eqnarray}

The conservation of lepton current requires that $q_\mu\pi^\munu(q)=0$ which fixes the Lorentz structure of $\pi^\munu$ as 
\begin{eqnarray}
\pi^\munu(q) = \FB{g^\munu-\frac{q^\mu q^\nu}{q^2}}\frac{1}{3}\pi^\alpha_{~\alpha}(q)~.
\end{eqnarray}
Substituting the above equation into Eq.~(\ref{eq.dpr.3}), and imposing the conservation of quark/antiquark current by means of $q^\mu\overline{\Pi}_\munu(q)=0$, we get,
\begin{eqnarray}
\text{DPR} = \FB{\frac{dN}{d^4qd^4x}} = \frac{4}{3(2\pi)^4q^4} \FB{\frac{1}{e^{\beta q^0}-1}}
\IM\overline{\Pi}^\mu_{~\mu}(q)~\IM\pi^\nu_{~\nu}(q)~.
\label{eq.dpr.4}
\end{eqnarray}
Let us now finally express the DPR in terms of electromagnetic spectral functions $\rho^{q\bar{q}}$ and $\rho^{l\bar{l}}$ as
\begin{eqnarray}
\text{DPR} = \FB{\frac{dN}{d^4qd^4x}} = \frac{4\alpha^2}{3\pi^2q^4} \FB{\frac{1}{e^{\beta q^0}-1}}
\rho^{q\bar{q}}(q)~\rho^{l\bar{l}}(q)~.
\label{eq.dpr.5}
\end{eqnarray}
where, $\rho^{q\bar{q}}$ is defined in Eq.~(\ref{eq.rho.def}) and $\rho^{l\bar{l}}$ defined as 
\begin{eqnarray}
\rho^{l\bar{l}}(q) = \frac{1}{4\pi\alpha}\IM\pi^\mu_{~\mu}(q)~.
\end{eqnarray}
Since we have already calculated $\rho^{q\bar{q}}$ in Eq.~(\ref{eq.rho.t}), it is trivial to write down the expression for $\rho^{l\bar{l}}$ by replacing $N_c\rightarrow1$, $T\rightarrow0$, $\sum_{f}^{}e_f^2\rightarrow e^2$ and $m_f\rightarrow m$ in Eq.~(\ref{eq.rho.t}). Considering the dileptons with physical momenta ($q^2>0$, $q^0>0$), only the Unitary-I cut contributes and we get,
\begin{eqnarray}
\rho^{l\bar{l}}(q) = \frac{1}{16\pi|\vec{q}|}\int_{\omega_-}^{\omega_+}d\omega_k U_1(\cos\theta=\cos\theta_0)
	\Theta\FB{q^0-\sqrt{\vec{q}^2+4m^2}}
\label{eq.rho.l}
\end{eqnarray}
where, 
\begin{eqnarray}
\omega_\pm &=& \frac{1}{2q^2}\TB{q^0q^2\pm|\vec{q}|\lambda^{1/2}\FB{q^2,m^2,m^2}} \label{eq.omegapm.lepton}\\
U_1 &=& \left.8\FB{2m^2-k^2-k\cdot q}\right|_{k^0=-\omega_k} \\
\cos\theta_0 &=& \FB{\frac{-2q^0\omega_k+q^2}{2|\vec{q}||\vec{k}|}} \label{eq.costheta0.lepton}
\end{eqnarray}
with $\omega_k=\sqrt{\vec{k}^2+m^2}$. Substituting Eq.~(\ref{eq.omegapm.lepton})-(\ref{eq.costheta0.lepton}) into Eq.~(\ref{eq.rho.l}) and performing the $d\omega_{k}$ integral, we arrive at,
\begin{eqnarray}
\rho^{l\bar{l}}(q) = \frac{q^2}{4\pi}\FB{1+\frac{2m^2}{q^2}}\FB{1-\frac{4m^2}{q^2}}^{1/2}\Theta\FB{q^2-4m^2}~.
\label{eq.rho.ll}
\end{eqnarray}
Substituting the $\rho^{l\bar{l}}(q)$ from the above equation into Eq.~(\ref{eq.dpr.5}) we obtain the final expression of DPR from QGP under \textit{zero external magnetic field} as
\begin{eqnarray}
\text{DPR} = \FB{\frac{dN}{d^4qd^4x}} = \frac{\alpha^2}{3\pi^3q^2}\FB{1+\frac{2m^2}{q^2}}\FB{1-\frac{4m^2}{q^2}}^{1/2} \FB{\frac{1}{e^{\beta q^0}-1}}\rho^{q\bar{q}}(q)\Theta\FB{q^2-4m^2}~.
\label{eq.dpr}
\end{eqnarray}
The presence of the unit step function in the above equation restricts the kinematic region where the DPR is non-zero. 


\section{Dilepton Production Rate under external magnetic field}\label{sec.dlr}
In order to calculate the DPR from a magnetized QGP medium, we start with Eq.~(\ref{eq.dpr.3}) 
\begin{eqnarray}
\text{DPR} = \FB{\frac{dN}{d^4qd^4x}} = \frac{4}{(2\pi)^4q^4} \FB{\frac{1}{e^{\beta q^0}-1}}
\IM\overline{\Pi}_\numu(q)~\IM\pi^\munu(q)~.
\label{eq.dpr.6}
\end{eqnarray}
where, the effect of external magnetic field will be entered through the photon polarization tensors $\overline{\Pi}_\munu(q)$ and $\overline{\pi}_\munu(q)$ calculated in presence of magnetic field. For the sake of simplicity in analytic calculation, we take the transverse momenta of the photon to be zero i.e. $q_\perp=0$ so that, in presence of external magnetic field, the Lorentz structure of $\pi^\munu(q_\parallel)$ will be
\begin{eqnarray}
\pi^\munu(q_\parallel) = \FB{g_\parallel^\munu-\frac{q_\parallel^\mu q_\parallel^\nu}{q_\parallel^2}}
\FB{\pi_\alphabeta g_\parallel^\alphabeta} + g_\perp^\munu \FB{\frac{1}{2}\pi_\alphabeta g_\perp^\alphabeta}~.
\end{eqnarray}
Substituting the above equation into Eq.~\ref{eq.dpr.6} and imposing the conservation of quark/antiquark current $\FB{q_\parallel^\mu\overline{\Pi}_\munu(q_\parallel)=0}$, we get,
\begin{eqnarray}
\text{DPR} = \FB{\frac{dN}{d^4qd^4x}} = \frac{4}{(2\pi)^4q_\parallel^4} \FB{\frac{1}{e^{\beta q^0}-1}}
\TB{g_\parallel^\munu g_\parallel^\alphabeta\IM\overline{\Pi}_\numu(q_\parallel)~\IM\pi_\alphabeta(q_\parallel)
+ \frac{1}{2}g_\perp^\munu g_\perp^\alphabeta\IM\overline{\Pi}_\numu(q_\parallel)~\IM\pi_\alphabeta(q_\parallel)}~.
\label{eq.dpr.7}
\end{eqnarray}
We define the longitudinal and transverse spectral functions as
\begin{eqnarray}
\rho_\parallel^{q\bar{q}}(q_\parallel) = \frac{1}{4\pi\alpha}g_\parallel^\munu\IM\overline{\Pi}_\numu(q_\parallel)
~~~~,~~~~
\rho_\perp^{q\bar{q}}(q_\parallel) = \frac{1}{4\pi\alpha}g_\perp^\munu\IM\overline{\Pi}_\numu(q_\parallel) 
\label{eq.rho.0} \\
\rho_\parallel^{l\bar{l}}(q_\parallel) = \frac{1}{4\pi\alpha}g_\parallel^\munu\IM\pi_\munu(q_\parallel)
~~~~,~~~~
\rho_\perp^{l\bar{l}}(q_\parallel) = \frac{1}{4\pi\alpha}g_\perp^\munu\IM\pi_\munu(q_\parallel)
\label{eq.rho.1}
\end{eqnarray}
in terms of which the DPR may be expressed as,
\begin{eqnarray}
\text{DPR} = \FB{\frac{dN}{d^4qd^4x}} = \frac{4\alpha^2}{\pi^2q_\parallel^4} \FB{\frac{1}{e^{\beta q^0}-1}}
\TB{\rho_\parallel^{q\bar{q}}(q_\parallel)\rho_\parallel^{l\bar{l}}(q_\parallel)
+ \frac{1}{2}\rho_\perp^{q\bar{q}}(q_\parallel)\rho_\perp^{l\bar{l}}(q_\parallel)}~.
\label{eq.dpr.8}
\end{eqnarray}

Next step is to calculate the longitudinal and transverse spectral functions defined in Eqs.~(\ref{eq.rho.0}) and (\ref{eq.rho.1}). We have already calculated   $\rho^{q\bar{q}}(q_\parallel)=\rho_\parallel^{q\bar{q}}(q_\parallel)+\rho_\perp^{q\bar{q}}(q_\parallel)$ in Eq.~(\ref{eq.rho.tb3}) which can be split into the transverse and longitudinal part. This is done by splitting Eq.~(\ref{eq.N.nl}) into a longitudinal and transverse part as
\begin{eqnarray}
N_{f,nl}(q_\parallel,k_\parallel)&=& (-1)^{n+l} \FB{\frac{|e_fB|}{\pi}}2\TB{4|e_fB|n\delta_{n-1}^{l-1}
	+\FB{\delta_n^l+\delta_{n-1}^{l-1}}m_f^2 + \FB{\delta_{n-1}^l+\delta_n^{l-1}}
	\FB{k_\parallel^2+k_\parallel\cdot q_\parallel-m_f^2}} \nn \\
&=& N_{f,nl,\parallel}(q_\parallel,k_\parallel) + N_{f,nl,\perp}(q_\parallel,k_\parallel)
\end{eqnarray} 
where, 
\begin{eqnarray}
N_{f,nl,\parallel}(q_\parallel,k_\parallel) &=& (-1)^{n+l} \FB{\frac{|e_fB|}{\pi}}2\TB{4|e_fB|n\delta_{n-1}^{l-1}
	+\FB{\delta_n^l+\delta_{n-1}^{l-1}}m_f^2\frac{}{}} \label{eq.N.nl.pll}\\
N_{f,nl,\perp}(q_\parallel,k_\parallel)&=& (-1)^{n+l} \FB{\frac{|e_fB|}{\pi}}2\TB{\FB{\delta_{n-1}^l+\delta_n^{l-1}}
	\FB{k_\parallel^2+k_\parallel\cdot q_\parallel-m_f^2}} ~.
\label{eq.N.nl.per}
\end{eqnarray} 
Therefore the expressions for $\rho_{\parallel,\perp}^{q\bar{q}}(q_\parallel)$ will be same as Eq.~(\ref{eq.rho.tb3}) except the fact that $N_{f,nl}(q,k)$ in Eq.~(\ref{eq.N.nl}) will be replaced by $N_{f,nl,\parallel}(q,k)$ or $N_{f,nl,\perp}(q,k)$ which are given in Eqs.~(\ref{eq.N.nl.pll}) and (\ref{eq.N.nl.per}).

The longitudinal and transverse spectral functions for leptonic case $\rho_{\parallel,\perp}^{l\bar{l}}(q_\parallel)$ can be obtained from $\rho_{\parallel,\perp}^{q\bar{q}}(q_\parallel)$ by replacing $T\rightarrow0$, $N_c\rightarrow1$ and $\sum_{f}^{}e_f^2\rightarrow e^2$. Considering the dileptons with physical momenta ($q^2>0$, $q^0>0$), only the Unitary-I cut contributes so that we get,
\begin{eqnarray}
\rho_{\parallel,\perp}^{l\bar{l}}(q_\parallel) = \sum_{n=0}^{\infty}~\sum_{l=(n-1)}^{(n+1)}\frac{\Theta\FB{q^0-\sqrt{q_z^2+(m_l+m_n)^2}}}
{4\lambda^{1/2}(q_\parallel^2,m_l^2,m_n^2)} ~
\sum_{\tilde{k}_z\in\tilde{k}_z^\pm}^{}
\TB{N^{nl}_{\parallel,\perp}(q_\parallel,\tilde{k}_z,k^0=-\tilde{\omega}_k^l)} 
\end{eqnarray}
where, $\tilde{k}_z^\pm = \frac{1}{2q_\parallel^2}\TB{-yq_z\pm|q^0|\lambda^{1/2}\FB{q_\parallel^2,m_l^2,m_n^2}}$, $y=(q_\parallel^2+m_l^2-m_n^2)$, $\tilde{\omega}_k^l = \sqrt{\tilde{k}_z^2+m_{l}^2}$ and $m_l=\sqrt{m^2+2l|eB|}$. $N^{nl}_{\parallel,\perp}(q_\parallel,k_\parallel)$ in the above equation are obtained from Eqs.~(\ref{eq.N.nl.pll}) and (\ref{eq.N.nl.per}) by replacing $e_f\rightarrow e$. Substituting $N^{nl}_{\parallel,\perp}(q_\parallel,k_\parallel)$ in the above equation and performing the sum over $\tilde{k}_z$, we get after some simplifications,
\begin{eqnarray}
\rho_{\parallel}^{l\bar{l}}(q_\parallel) = \sum_{n=0}^{\infty}~\sum_{l=(n-1)}^{(n+1)}\frac{\Theta\FB{q_\parallel^2-(m_l+m_n)^2}}
{\lambda^{1/2}(q_\parallel^2,m_l^2,m_n^2)} ~
\frac{|eB|}{\pi}\TB{4|eB|n\delta_{n-1}^{l-1}+m^2\FB{\delta_n^l+\delta_{n-1}^{l-1}}\frac{}{}} 
\label{eq.rho.pll}\\
\rho_{\perp}^{l\bar{l}}(q_\parallel) = \sum_{n=0}^{\infty}~\sum_{l=(n-1)}^{(n+1)}\frac{\Theta\FB{q_\parallel^2-(m_l+m_n)^2}}
{\lambda^{1/2}(q_\parallel^2,m_l^2,m_n^2)}\SB{\frac{1}{2}\FB{q_\parallel^2+m_l^2-m_n^2}-2l|eB|}
\frac{|eB|}{\pi}\FB{\delta_n^{l-1}+\delta_{n-1}^l}
\label{eq.rho.per}
\end{eqnarray}
For a consistency check, let us consider LLL approximation for which we have,
\begin{eqnarray}
\rho_{\parallel,\text{LLL}}^{l\bar{l}}(q_\parallel)&=&\frac{|eB|}{\pi}\frac{m^2}{q_\parallel^2}\FB{1-\frac{4m^2}{q_\parallel^2}}^{-1/2}
\Theta\FB{q_\parallel^2-4m_l^2} \\
\rho_{\perp,\text{LLL}}^{l\bar{l}}(q_\parallel) &=& 0
\end{eqnarray}
which agrees with the the expression in Ref.~\cite{Bandyopadhyay:2016fyd} where the authors have obtained the leptonic contribution by using the spin sum over leptonic spinors in presence of external magnetic field.

The presence of the step functions in Eqs.~(\ref{eq.rho.pll}) and (\ref{eq.rho.per}) restricts the kinematic domain for the non-zero DPR in presence of external magnetic field. Let us consider the case when the lepton mass is neglected. In that case, the contribution from the LLL vanishes and thus the kinematic region for non-zero DPR starts from $q_\parallel^2 > 2|eB|$. It will have consequence that in presence of strong enough external magnetic field, the low-invariant mass dileptons ($q_\parallel^2<2|eB|$) can not be produced. We will discuss this in detail in the next section.

\section{Results and discussions}\label{sec.numerical}
\begin{figure}[h]
	\begin{center}
		\includegraphics[angle=-90, scale=0.35]{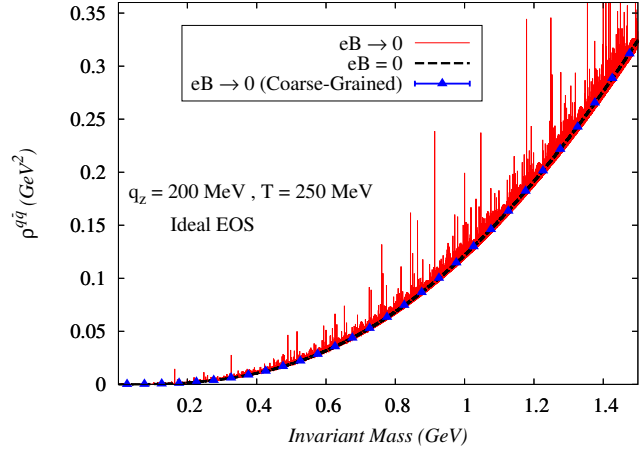}
	\end{center}
\caption{Electromagnetic spectral function due to quark loop as a function of photon invariant mass at $T$ = 250 MeV, $q_z$ = 200 MeV for ideal EOS. Results for $eB=0$ has been compared among that of $eB\rightarrow0$ and the coarse-grained $eB\rightarrow0$. }
	\label{fig.spectra.2}
\end{figure}

\begin{figure}[h]
	\begin{center}
		\includegraphics[angle=-90, scale=0.35]{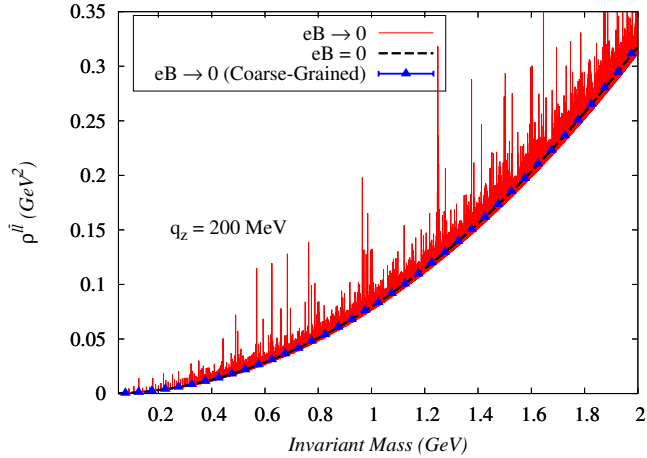}
	\end{center}
	\caption{Electromagnetic spectral function due to lepton loop as a function of photon invariant mass at $q_z$ = 200 MeV. Results for $eB=0$ has been compared among that of $eB\rightarrow0$ and the coarse-grained $eB\rightarrow0$.  }
	\label{fig.spectra.6}
\end{figure}

The numerical results presented in this work are obtained considering (2+1) flavours of the quarks where the masses of up and down quarks are taken zero ($m_u=m_d=0$) and the strange quark mass is $m_s=100$ MeV. 
It is to be noted that, even at very high temperature ($\sim 250$ MeV), the quarks might still have the effects from
the chiral dynamics due to the crossover behavior of the QCD transition from hadron to the QGP in terms of the temperature  dependence in their thermal masses~\cite{Schaefer:1998wd}. Similar studies incorporating the temperature dependent quark mass for the calculation of DPR can be found in Refs.~\cite{Peshier:1999dt,Mustafa:1999dt,Greiner:2010zg}. In addition to the temperature dependence, 
the quark mass can also depend on the quark virtuality as discussed in Refs.~\cite{Dorokhov:2001wx,Roberts:1994dr,Schafer:1996wv}. In fact, the quark mass asymptotically reaches to the current quark mass values at higher Euclidean momenta of quarks corresponding to the perturbative regime. Thus the effect of momentum dependence of quark mass at the high temperature  is expected to be small. 
In this work, we have not considered these complications and used constant quark mass in the numerical calculations. In our approach, the effective fugacities capture all the medium effects and contribute to the screening masses of the quarks and antiquarks through the effective couplings.
The lepton mass is also considered as zero ($m=0$). For convergent results at non-zero external magnetic field cases, we have considered upto 1000 Landau levels in the numerical calculations. However, for the taking $eB\rightarrow 0$ limit numerically in Figs.~\ref{fig.spectra.2} and \ref{fig.spectra.6}, upto 10000 Landau levels are considered.

\begin{figure}[h]
\begin{center}
\includegraphics[angle=-90, scale=0.35]{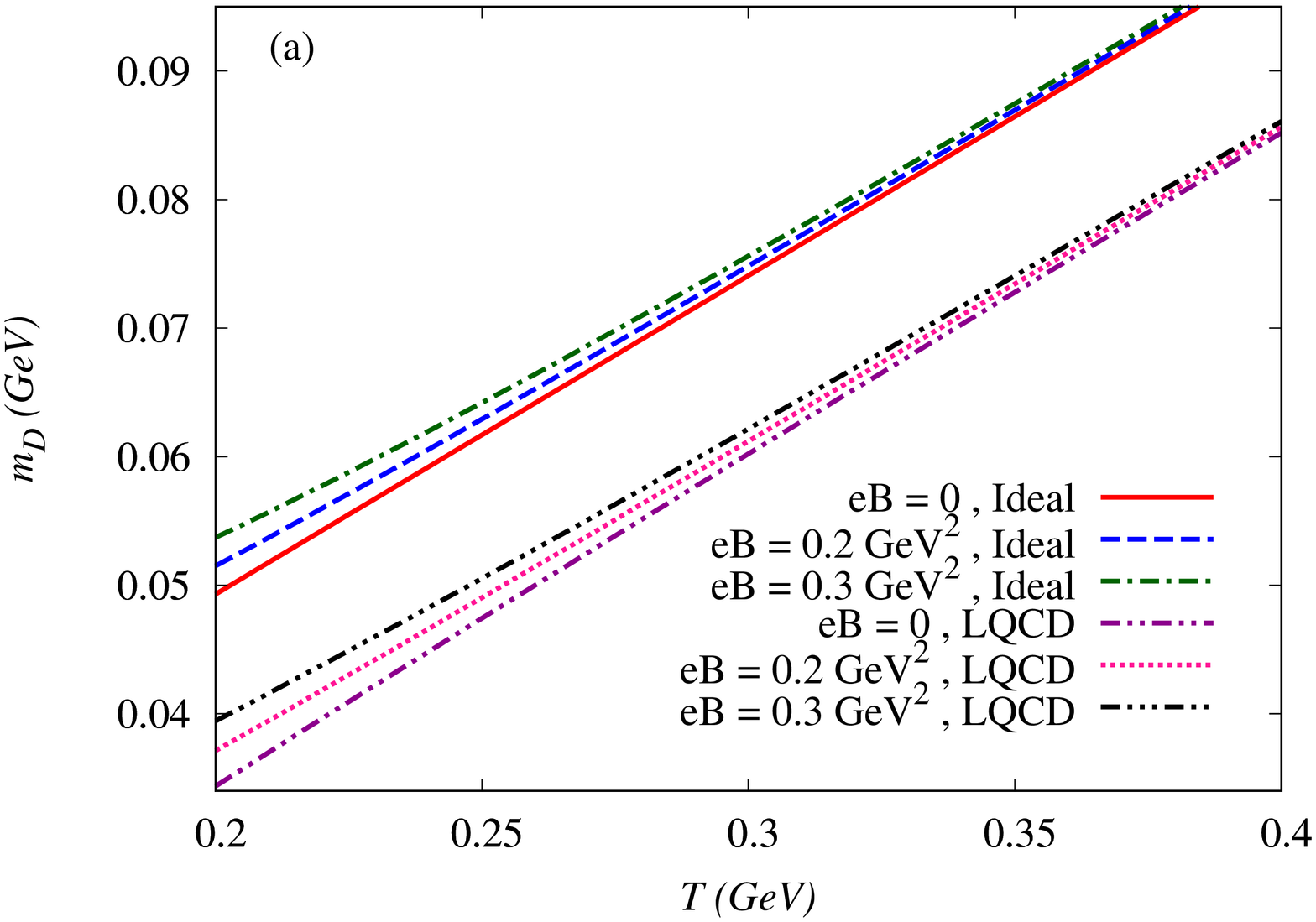}
\includegraphics[angle=-90, scale=0.35]{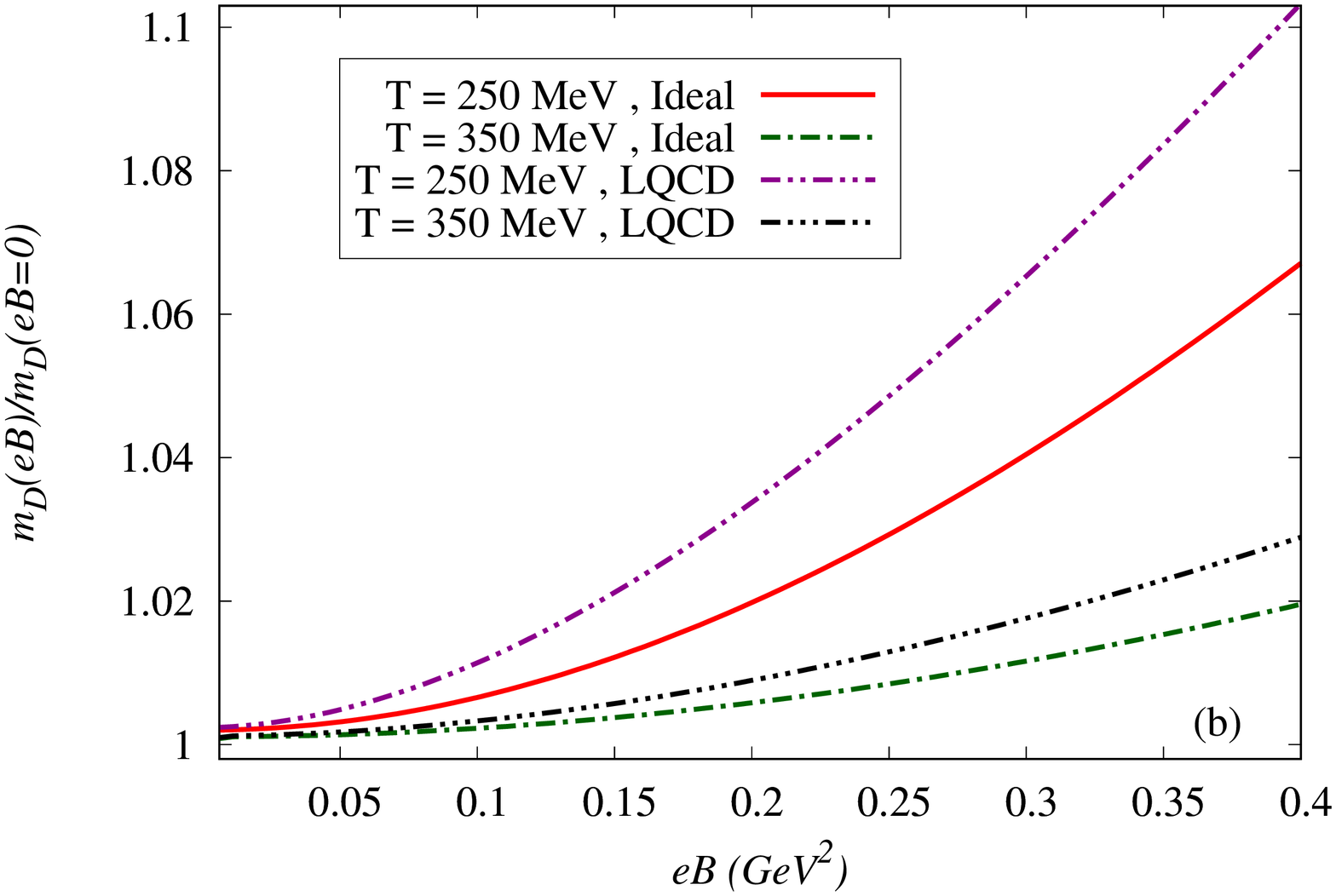}
\end{center}
\caption{(a) Debye Mass as a function of temperature at different values of external magnetic field ($eB$ = 0, 0.2 and 0.3 GeV$^2$) for different EOSs. (b) The ratio of Debye mass at finite $eB$ to the Debye mass at zero $eB$ as a function of $eB$ at two different temperatures ($T$=250 and 350 MeV) for different EOSs. The ``Ideal EOS" corresponds to $z_q=1$ and ``LQCD EOS" corresponds to $z_q=z_q(T)<1$ which is obtained from Ref.~\cite{Bazavov:2009zn,Bazavov:2014pvz}.}
\label{fig.debye}
\end{figure}

Let us start this section by presenting the variation of Debye mass as a function of temperature in Fig~\ref{fig.debye}(a). At $eB=0$, the $m_D$ for the ideal EOS increases monotonically with the increase in $T$ which is obvious from Eq.~(\ref{eq.md}) in which the thermal distribution functions increases with the increase in temperature. At very high temperature, for which the strange quark mass can be neglected with respect to $T$, we proceed towards the linear relationship of $m_D$ with $T$ as given in Eq.~(\ref{eq.md.ideal}). For the LQCD EOS $m_D$ shows similar variation with $T$ as that of the ideal one. However the $m_D$ at a particular temperature has a lower value for LQCD EOS as compared to the ideal one. This is due to fact that quasi quarks distribution functions contain the $z_q(T)$ which is less than unity. The effect of the strong interactions encoded in the $z_q(T)$ thus have significant effect on $m_D$. 
With the increase in $eB$, the $m_D$ also increases but by a small amount for the two EOSs. To see the effect of $eB$ more clearly, we have plotted the ratio $m_D(eB)/m_D(eB=0)$ as a function of $eB$ in Fig.~\ref{fig.debye}(b). We find, the effect of $eB$ is more in case of LQCD EOSs with respect to the ideal EOS. However, for a signifiant high magnetic field $eB\sim0.4$ GeV$^2$, the Debye mass enhances by less than 10$\%$.

The Debye screening mass enhancement in presence of external magnetic field is in accordance to the finding  in Ref.~\cite{Bonati:2017uvz} using Lattice QCD simulations and in Refs.\cite{Alexandre:2000jc,Bandyopadhyay:2017cle,Bandyopadhyay:2016fyd}~ using perturbative calculation. This effect in certain range of temperatures might be related to the chiral symmetry breaking in terms of magnetic catalysis as argued in Refs.~\cite{Bandyopadhyay:2016fyd,Alexandre:2000jc}. 
However, to make a concrete statement, we require to know the role of external electromagnetic field on plasma relaxation processes in QGP which is beyond the scope of present study.

\begin{figure}[h]
	\begin{center}
		\includegraphics[angle=-90, scale=0.35]{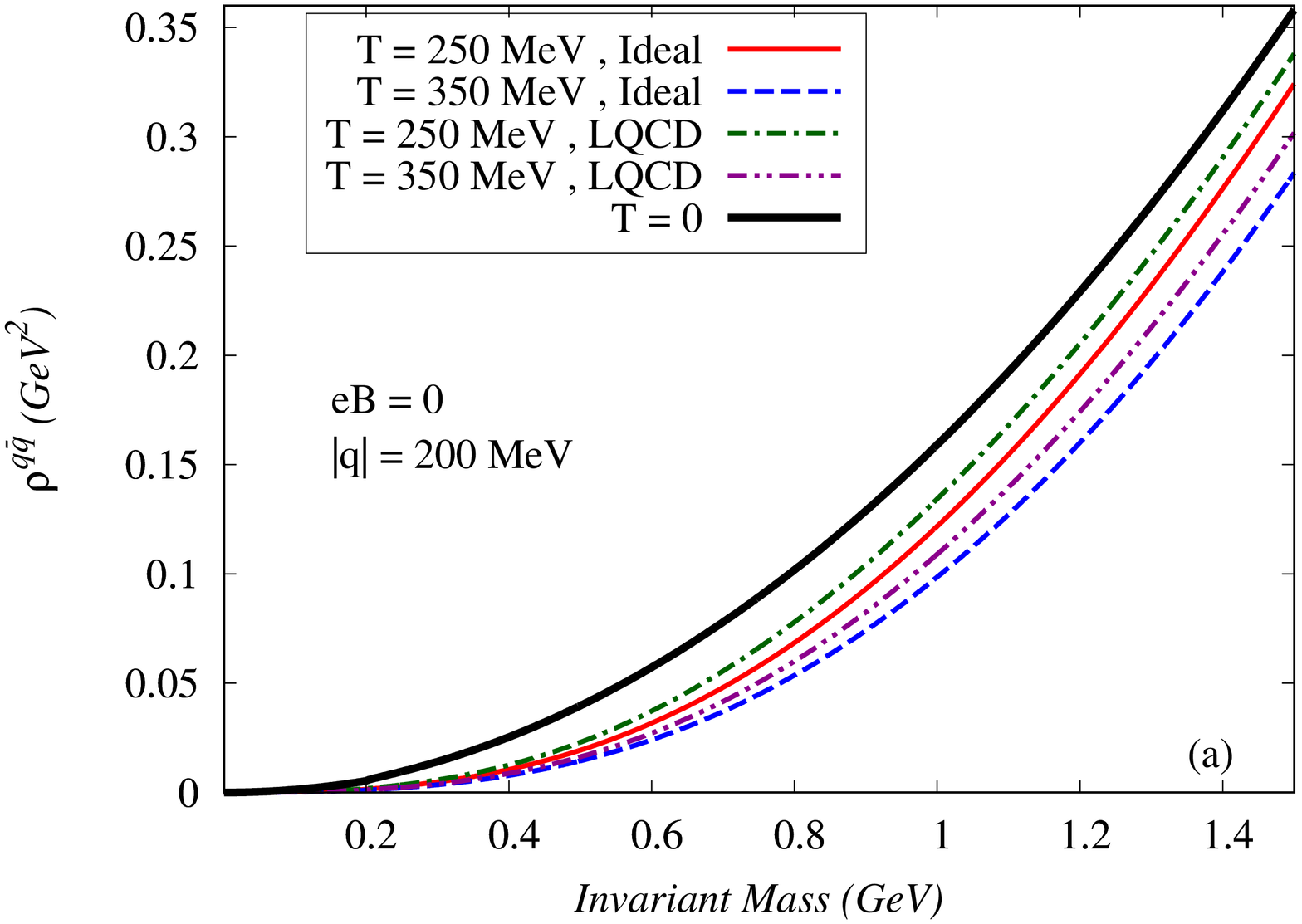}
		\includegraphics[angle=-90, scale=0.35]{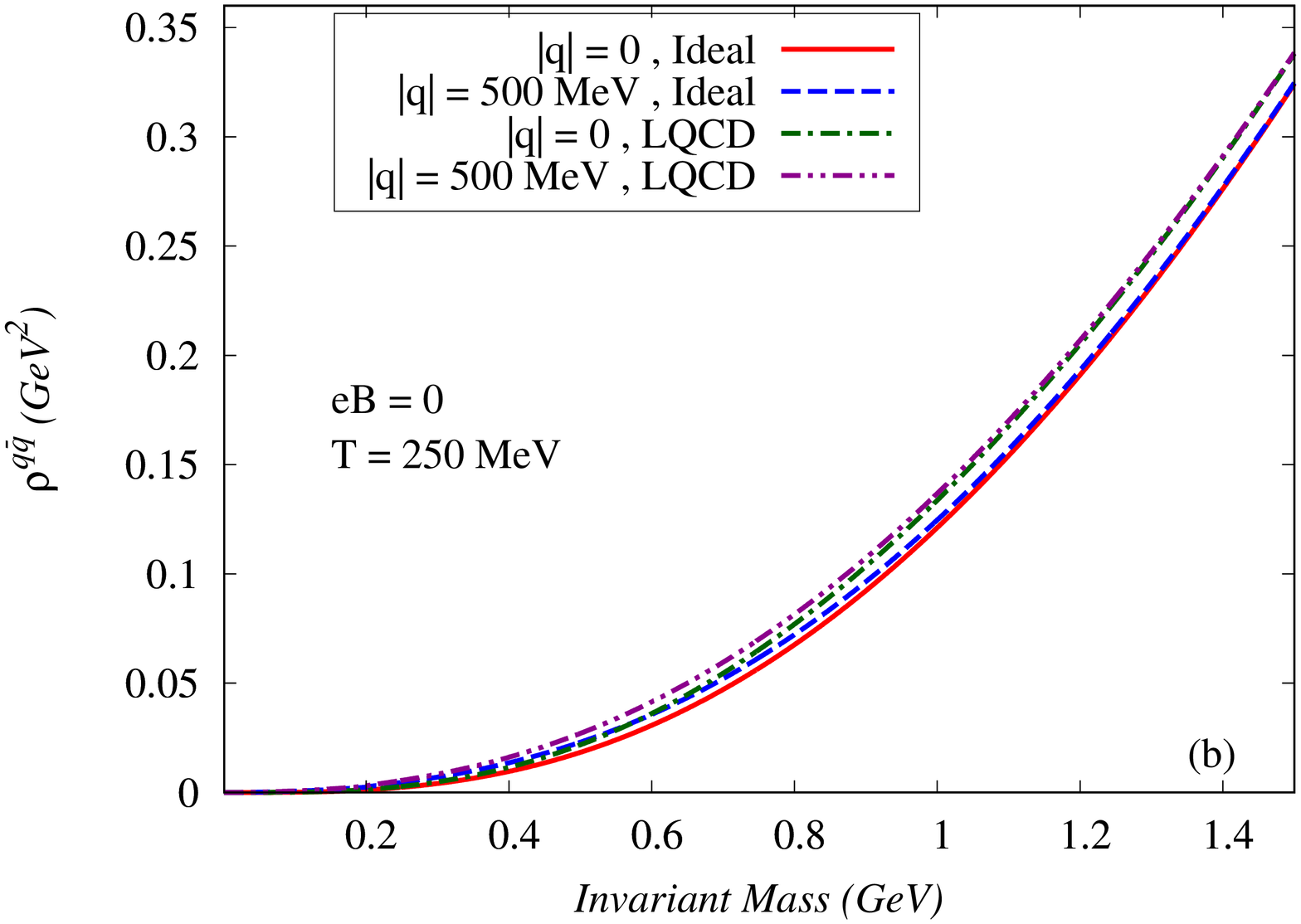}		
	\end{center}
\caption{Electromagnetic spectral function due to quark loop for different EOSs as a function of photon invariant mass at zero external magnetic field (a) with photon three momentum $\vec{q}$ = 200 MeV at two different temperatures (250 and 350 MeV) (b) atconstant temperature (250 MeV) and at two different photon three momentum $\vec{q}$ (0 and 500 MeV). The Vacuum spectral function (T = 0) is also shown in sub-figure (a) for comparison.}
	\label{fig.spectra.1}
\end{figure}
Next we show the results of electromagnetic spectral functions due to quark loop in the magnetized hot QCD medium. Let us first consider the \textit{zero magnetic field} case in Fig~\ref{fig.spectra.1}. We have plotted the $\rho^{q\bar{q}}(q)$ as a function of invariant mass $\sqrt{q^2}$ at $|\vec{q}|=$ 200 MeV and at two different temperatures (250 and 350 MeV) for different EOSs in Fig~(\ref{fig.spectra.1})(a). The zero temperature case is also shown for comparison. As we have discussed in the previous section, in absence of $eB$, only the Unitary-I cut contributes to the spectral function which starts from $q^2 \ge 4m_f^2$. Since we have taken $m_u=m_d=0$, the spectral function is non zero at $q^2>0$. With the increase in invariant mass, $\rho^{q\bar{q}}(q)$ increases monotonically due to the increase in the availability of the phase space. Additionally, with the increase in $T$, the spectral function decreases. It can be understood from Eq.~(\ref{eq.rho.t}) where, the quark distribution functions increases with increase in $T$ in turn reduces the overall factor $\tanh\FB{\frac{\beta q^0}{2}}\SB{1-f(\omega_k^f)-f(\omega_p^f)+2f(\omega_k^f)f(\omega_p^f)}$ in the Unitary cut restricting the phase space availability of the quarks. The effect of inclusion of effective fugacity in quasi quark distribution function has an opposite effect with respect to the increase in temperature. Because of $z_q(T) <1$, the phase space is more in the LQCD EOS case as compared to the ideal one. In Fig~(\ref{fig.spectra.1})(b) we have shown $\rho^{q\bar{q}}(q)$ as a function of invariant mass at a constant temperature $T=250$ MeV and at two different values of photon three momentum ($|\vec{q}|=0$ and 500 MeV) for the different EOSs. It is observed that, the effect of increase of $|\vec{q}|$ on $\rho^{q\bar{q}}(q)$ is very small and it only affects at the low invariant mass region.
\begin{figure}[h]
	\begin{center}
		\includegraphics[angle=-90, scale=0.35]{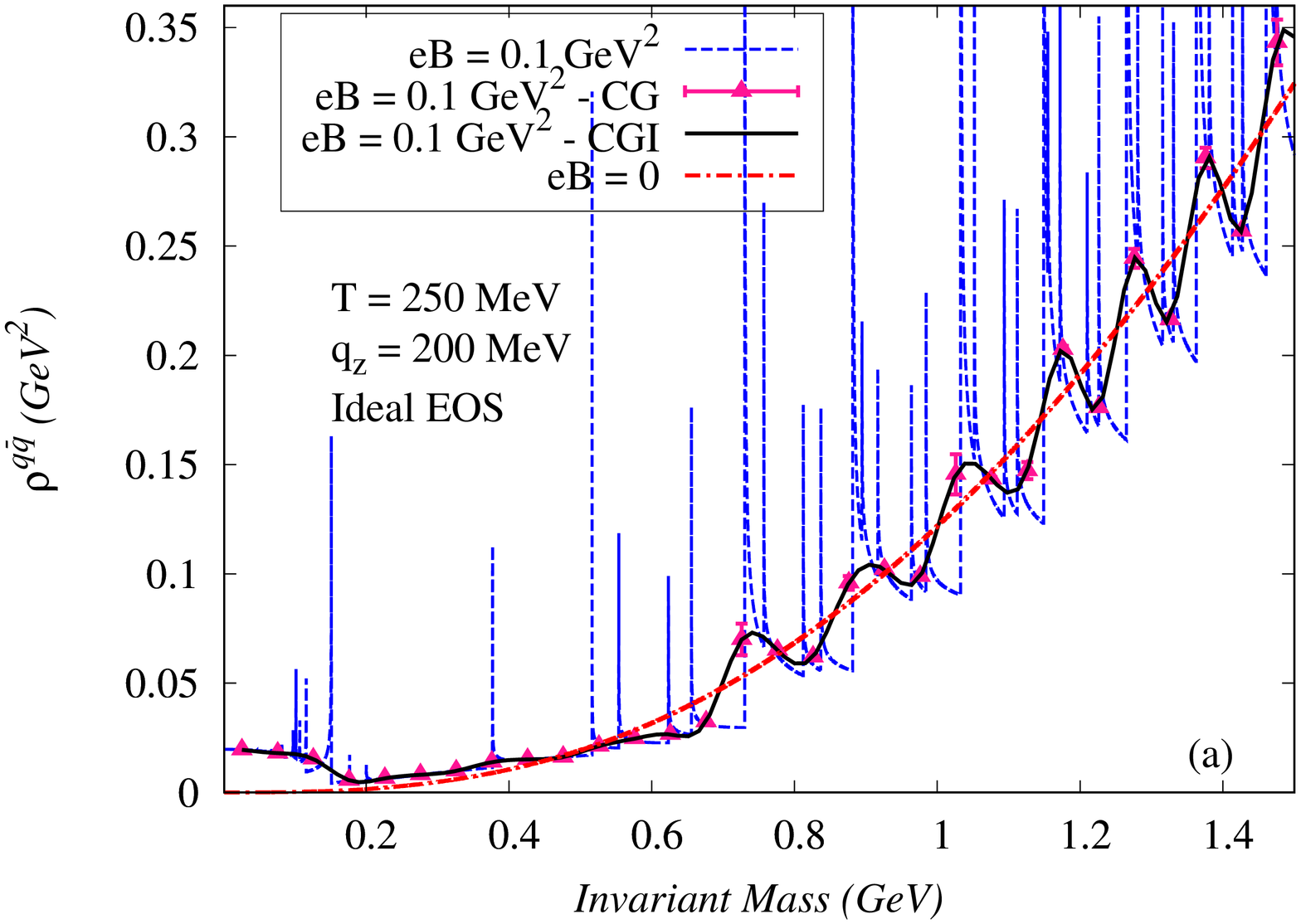}
		\includegraphics[angle=-90, scale=0.35]{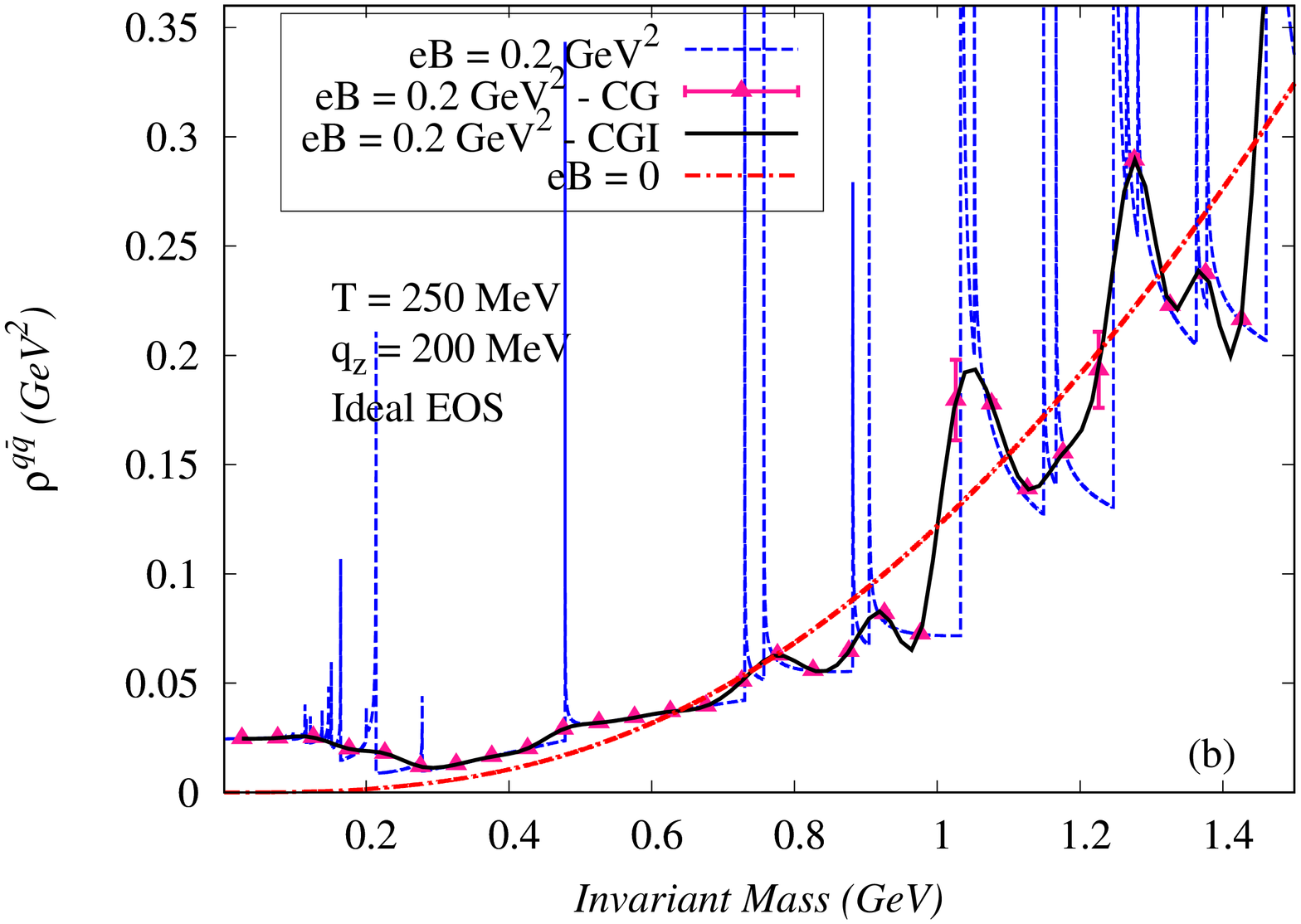}		
	\end{center}
\caption{Electromagnetic spectral function due to quark loop as a function of photon invariant mass at $T$ = 250 MeV, $q_z$ = 200 MeV for ideal EOS. Results for $eB=0$ has been presented along wih (a) $eB=0.1$ and (b) $eB=0.2$ GeV$^2$. The coarse-grained (CG) as well as coarse-grained interpolated (CGI) results for non zero $eB$ are also shown for comparison.}
	\label{fig.spectra.3}
\end{figure}

We now turn on the \textit{external magnetic field}. For a consistency check, we have first taken the $eB\rightarrow 0$ limit numerically to Eq.~(\ref{eq.rho.tb3}) and compared with $\rho^{q\bar{q}}$ calculated from Eq.~(\ref{eq.rho.t}) in Fig~\ref{fig.spectra.2}. The results are presented for ideal EOS at $T=$ 250 MeV and $q_z=$ 200 MeV. We found a large number of spikes infinitesimally spaced from each other covering the whole invariant mass axis. The $eB=0$ graph is analytic having no such spikes and interestingly it goes on average through the $eB\rightarrow 0$ graph. The appearance of these spikes are due the ``Threshold Singularities" in each Landau levels as can be seen from Eq.~(\ref{eq.rho.tb3}), where the K\"all\'en function in the denominator blows up in each threshold defined in terms of the step functions therein. This type of threshold singularities are observed in other works as well \cite{Bandyopadhyay:2016fyd,Chakraborty:2017vvg}. In order to extract finite and physical results for the spectral function we have introduced the concept of Ehrenfest's coarse-graining~\cite{Gorban,Ehrenfest} in which we have discretized the whole invariant mass region in small bins followed by performing bin averages. Thus, in other words, $\rho^{q\bar{q}}(q_\parallel)$ at a given $\sqrt{q_\parallel^2}$ is approximated by its average over the neighbourhood around that point. We calculated the coarse-grained (CG) spectral functions and show in Fig~\ref{fig.spectra.2}. It can be noticed that, the CG $\rho^{q\bar{q}}(q)$ for $eB\rightarrow 0$ exactly reproduce the $eB=0$ case.

We now increase the $eB$ and shown the spectral function due to quark loop for finite values of magnetic field ($eB=$ 0.1 and 0.2 GeV$^2$) in Fig.~\ref{fig.spectra.3}(a) and (b) respectively. The EOS, $T$ and $q_z$ are taken same as Fig.~\ref{fig.spectra.2}. The $eB=0$ case is also shown for comparison. As discussed in the previous section, in this case the Unitary-I cut begins from $q_\parallel^2\ge4m_f^2$. Since we have taken $m_u=m_d=0$, the threshold of the Unitary-I cut for these two massless flavours is $q_\parallel^2 \ge0$ whereas the same is $q_\parallel^2 \ge 4m_s^2$ for the strange quark. It is to be noted from the discussions in Appendix~\ref{app.kinematic.domain} that, Unitary-I cut threshold comes from the LLL. However for the massless quark flavours the contribution to the spectral function from LLL vanishes as evident from Eq.~(\ref{eq.N.nl}). Hence for the massless quark flavours, the threshold of the Unitary-I cut will be $q_\parallel^2 > 2|e_fB|$ which comes from the next to LLL. Thus when all the quark flavours are summed up, the threshold of the Unitary-I cut will become $q_\parallel^2 > \text{min}\FB{2\times\frac{1}{3}|eB|,4m_s^2}$. The spectral functions sufferer lots of ``Threshold Singularity" which are now separated from each other by finite value of invariant mass (unlike Fig.~\ref{fig.spectra.2} where they were infinitesimally spaced). With the increase in $eB$ the spacing among these spikes increases as can be seen from Eq.~(\ref{eq.rho.tb3}). Moreover the spectral functions for non-zero $eB$ perform oscillatory behaviours about the $eB=0$ graph at higher invariant mass region. This is more clearly visible when we plot the CG spectral functions. The CG spectral functions at the discrete bin points are used to obtain an interpolated graph which we call CG Interpolated (CGI) spectral function. The oscillatory behaviours of the spectral function can be best observed in the CGI graphs. This oscillation frequency is more at the lower $eB$ (0.1 GeV$^2$) as compared to the higher $eB$ (0.2 GeV$^2$) but the amplitude of oscillation is just the opposite (low for lower $eB$ and high for higher $eB$). This is consistent with Fig.~\ref{fig.spectra.1} where at a limiting value of $eB\rightarrow 0$, the oscillation frequency becomes infinite and amplitude becomes zero reproducing the $eB=0$ graph. It is important to mention that, in the case of non-zero $eB$, the Landau cuts also contribute (as discussed in the previous section) and in fact the lower invariant mass region is dominated by this Landau terms which will be clear in next paragraph.

\begin{figure}[h]
	\begin{center}
		\includegraphics[angle=-90, scale=0.35]{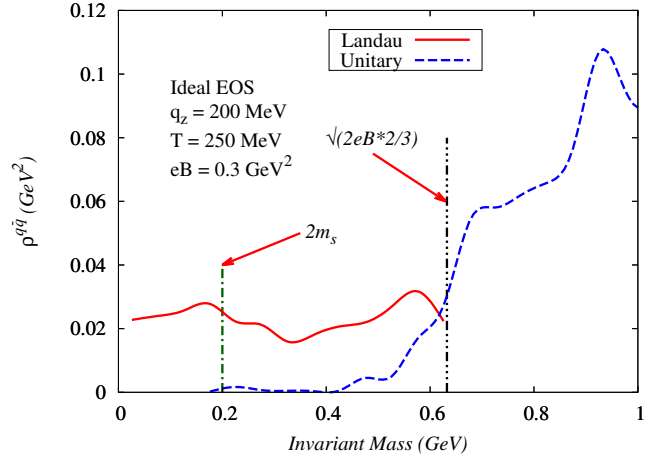}
	\end{center}
\caption{The Landau and Unitary contribution to the coarse-grained interpolated electromagnetic spectral function due to quark loop as a function of photon invariant mass at $T$ = 250 MeV, $q_z$ = 200 MeV and $eB=$ 0.3 GeV$^2$ for ideal EOS. The vertical black line correspond to the Landau-cut threshold which is $\sqrt{2eB\times2/3}$ (due to the up quark). The vertical green line correspond to the Unitary-cut threshold which is $2m_s$ (due to the strange quark).}
	\label{fig.spectra.4}
\end{figure}

In order to see significance of the Landau terms, we have compared the contribution to $\rho^{q\bar{q}}(q)$ arising from the Unitary and Landau cuts separately in Fig~\ref{fig.spectra.4}. The result is obtained for ideal EOS with $T=$ 250 MeV, $q_z=$ 200 MeV and $eB=$ 0.3 GeV$^2$. As can be seen from the figure, low invariant mass region is dominated by the Landau cut contribution where as the high invariant mass region is dominated by Unitary cut contribution. Moreover, the Landau cut is extended only up to $\sqrt{q_\parallel^2} \le \sqrt{2|Q_f|eB}$ with the maximum $Q_f=2/3$ for the up quark as can be understood from the previous section. This threshold of Landau cut is shown by black horizontal line in the figure. As discussed earlier, the Unitary cut threshold in this case is  $\sqrt{q_\parallel^2} \ge\text{min}\FB{\sqrt{2\times\frac{1}{3}|eB|},2m_s}=2m_s$ (since $eB=0.3$ GeV$^2$, $m_s=100$ MeV) which is shown by green vertical line in the figure.

\begin{figure}[h]
	\begin{center}
		\includegraphics[angle=-90, scale=0.35]{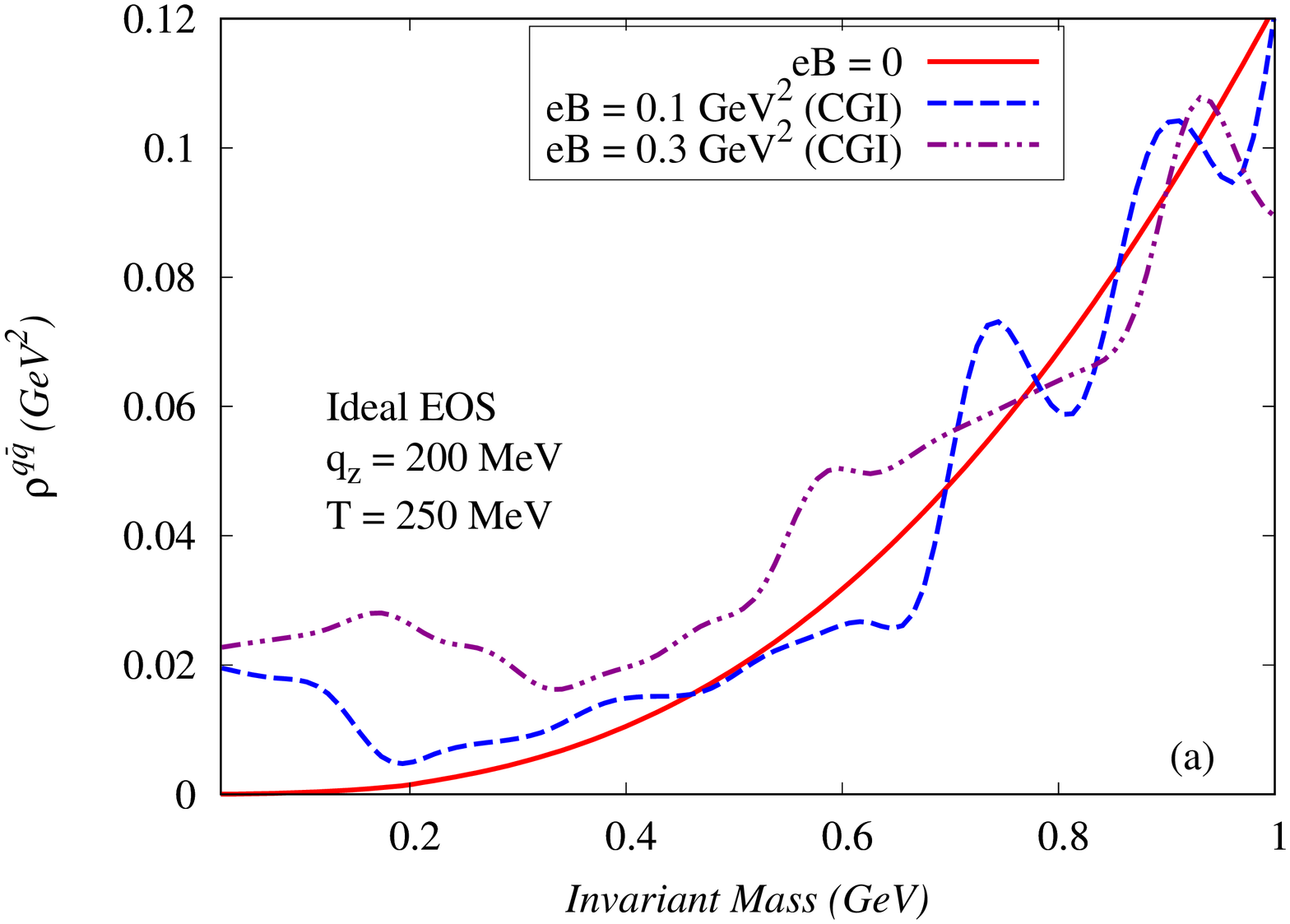}
		\includegraphics[angle=-90, scale=0.35]{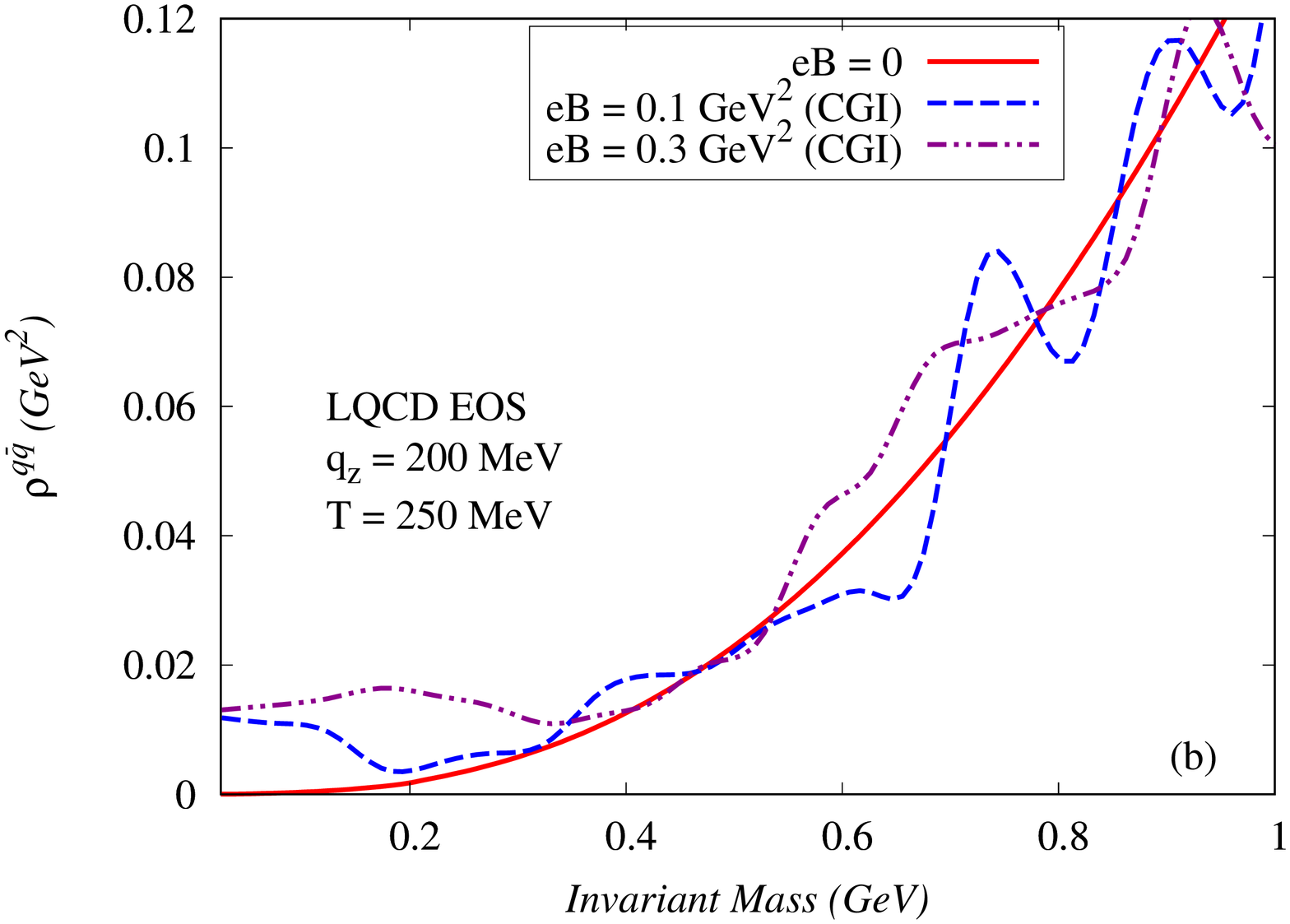}
	\end{center}
\caption{The coarse-grained interpolated electromagnetic spectral function due to quark loop as a function of photon invariant mass at $T$ = 250 MeV, $q_z$ = 200 MeV and three different vacules of $eB$ (0, 0.1 and 0.3 GeV$^2$) for (a) ideal and (b) LQCD EOS.}
	\label{fig.spectra.5}
\end{figure}

In Fig~\ref{fig.spectra.5}(a) we have shown CGI $\rho^{q\bar{q}}(q)$ as a function of invariant mass for three different values of $eB$ (0, 0.1 and 0.3 GeV$^2$) at $T=$ 250 MeV and $q_z=$ 200 MeV for Ideal EOS. We find significance enhancement of $\rho^{q\bar{q}}(q)$ with the increase in $eB$ at the low invariant mass regions where as at the higher values of invariant mass, the spectral function is oscillatory about the $eB=0$ graph. Analogous results for the LQCD EOS is presented in Figs~\ref{fig.spectra.5}(b).

\begin{figure}[h]
	\begin{center}
		\includegraphics[angle=-90, scale=0.35]{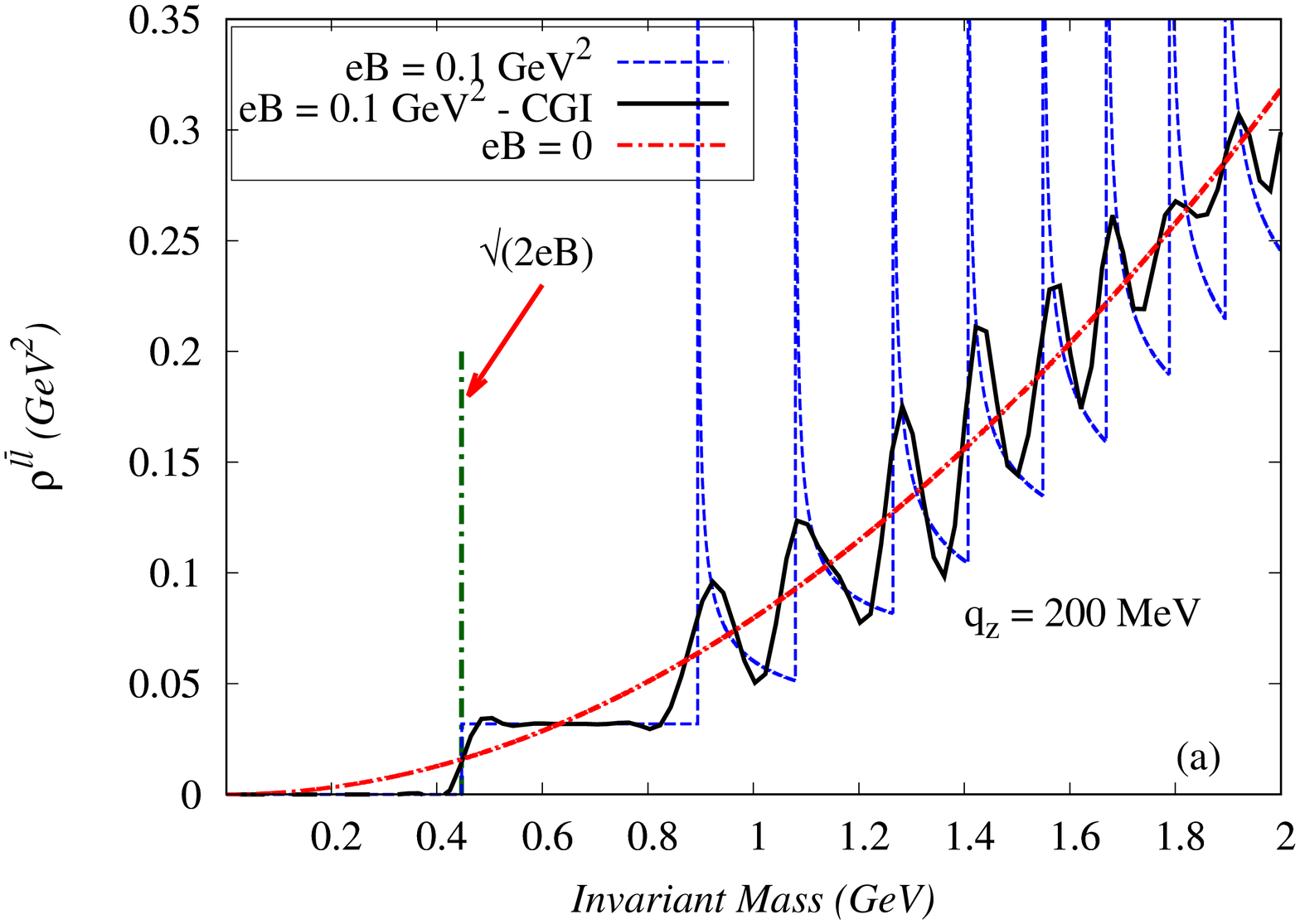}
		\includegraphics[angle=-90, scale=0.35]{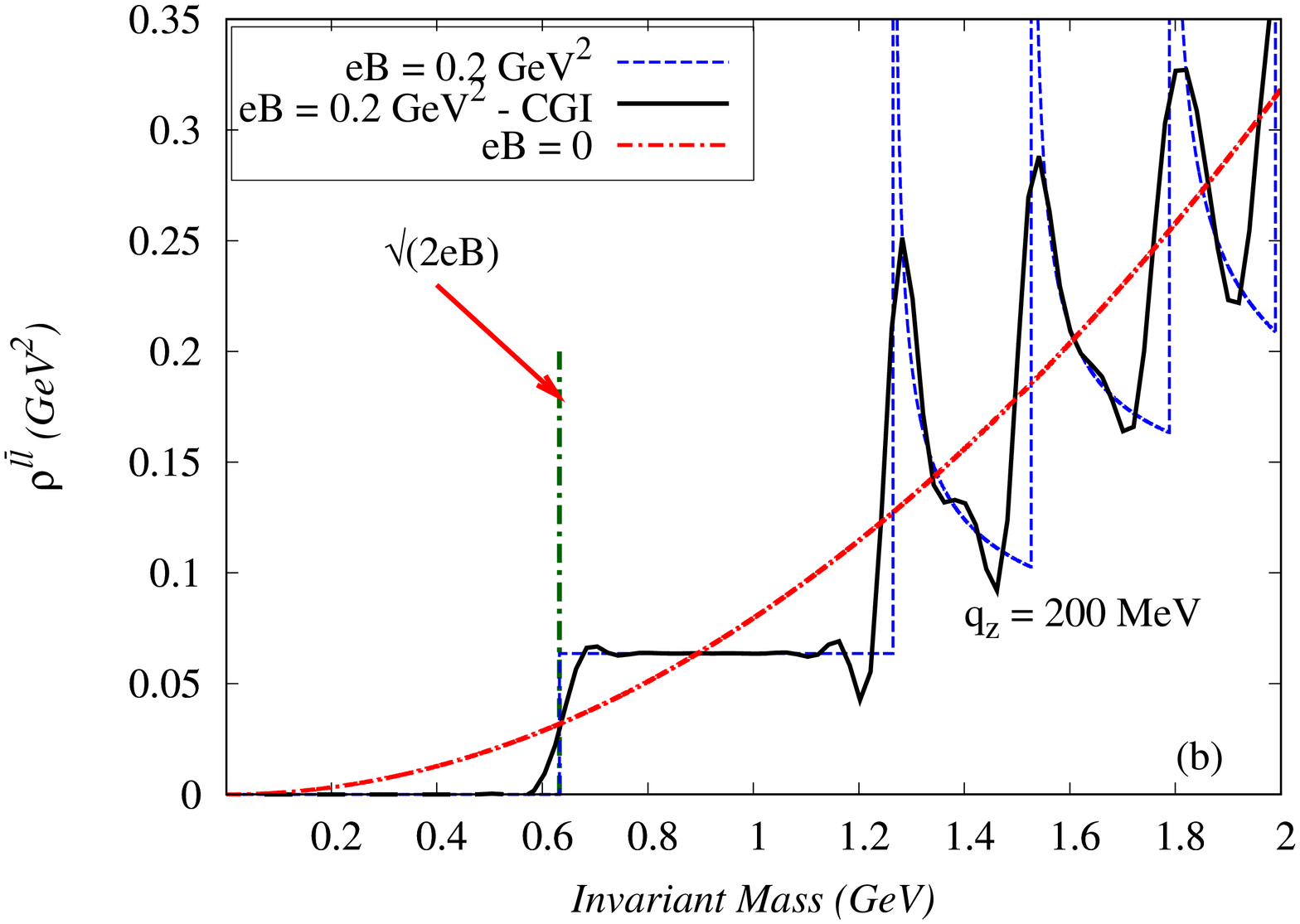}		
	\end{center}
\caption{Electromagnetic spectral function due to lepton loop as a function of photon invariant mass at $q_z$ = 200 MeV. Results for $eB=0$ has been presented along wih (a) $eB=0.1$ and (b) $eB=0.2$ GeV$^2$. The coarse-grained (CG) as well as coarse-grained interpolated (CGI) results for non zero $eB$ are also shown for comparison. The vertical green lines correspond to the Unitary cut threshold $\sqrt{q_\parallel^2}=\sqrt{2eB}$ for the massless leptons in presence of external magnetic field}
	\label{fig.spectra.7}
\end{figure}

Let us now turn our attention for the results of electromagnetic spectral function due to lepton loop $\rho^{l\bar{l}}=\rho_\parallel^{l\bar{l}}+\rho_\perp^{l\bar{l}}$. For a consistency check, we have first taken the $eB\rightarrow 0$ limit numerically from Eqs.~(\ref{eq.rho.pll}) and (\ref{eq.rho.per}) and compared with $\rho^{l\bar{l}}$ calculated from Eq.~(\ref{eq.rho.ll}) in Fig~\ref{fig.spectra.6}. The results are presented for $q_z=$ 200 MeV. Analogous to $\rho^{q\bar{q}}$, we found a large number of infinitesimally spaced spikes covering the whole invariant mass axis. The $eB=0$ graph goes on average through the $eB\rightarrow 0$ graph. The appearance of these spikes are due to the ``Threshold Singularities" in each Landau levels as can be seen from Eqs.~(\ref{eq.rho.pll}) and (\ref{eq.rho.per}) where the K\'allen lambda function in the denominators blows up in each threshold defined in terms of the step functions therein. The CG spectral function in this case also reproduces the $eB=0$ graph exactly. It is to be noted that, in this case we do not have any Landau cut contribution since the temperature is zero.

Next, in Fig.~\ref{fig.spectra.7}(a) and (b), $\rho^{l\bar{l}}$ is shown at $eB=0.1$ and 02 GeV$^2$ respectively for $q_z=200$ MeV. In this case the threshold of the Unitary cut begins from $\sqrt{q_\parallel}>\sqrt{2eB}$ as discussed in the last paragraph of the previous section. The spectral function shows oscillatory behaviour about the $eB=0$ graph at non-zero $eB$ and the oscillation frequency (amplitude) decreases (increases) with the increase of external magnetic field.

\begin{figure}[h]
	\begin{center}
		\includegraphics[angle=-90, scale=0.35]{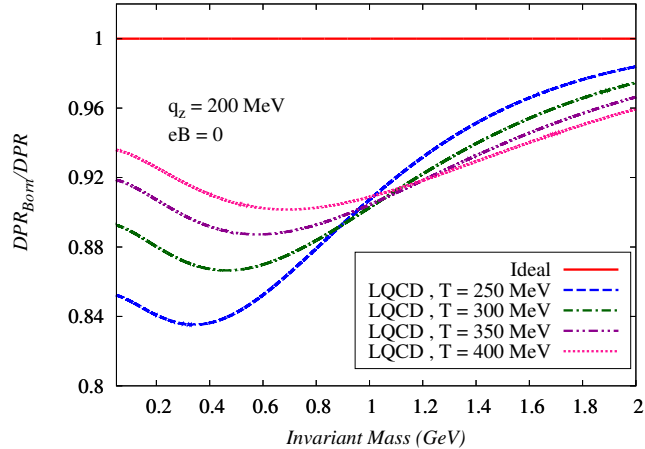}
	\end{center}
\caption{The ratio of Born rate to DPR as a function of dilepton invariant mass at zero external magnetic field for $q_z$ = 200 MeV. Results for Ideal EOS has been compared among LQCD EOS at four different temperatures (250, 300, 350 and 400 MeV). }
	\label{fig.dlr.1}
\end{figure}

Next, we investigate how the DPR from hot QCD medium gets modified due the external magnetic field as well as the EOS effects. Let us first consider the $eB=0$ case. We have plotted the ratio DPR$_\text{Born}$/DPR as a function of invariant mass of the dilepton in Fig~\ref{fig.dlr.1} where, DPR$_\text{Born}$ is the Born rate for the dilepton production given by~\cite{Sadooghi:2016jyf} (for lepton mass $m=0$)
\begin{eqnarray}
\text{DPR}_\text{Born} = \frac{\alpha^2N_cT}{6\pi^4|\vec{q}|}\FB{\frac{1}{e^{\beta q^0}-1}}
\sum_{f}^{} Q_f^2\FB{1+\frac{2m_f^2}{q^2}}
\ln\TB{\frac{\cosh\FB{\frac{q^0+R_f|\vec{q}|}{4T}}}{\cosh\FB{\frac{q^0-R_f|\vec{q}|}{4T}}}}\Theta\FB{q^2-4m_f^2}~.
\end{eqnarray}
where $R_f=\FB{1-\frac{4m_f^2}{q^2}}^{1/2}$. 
The results are obtained at four different temperature (250, 300, 350 and 400 MeV) and at $q_z=$ 200 MeV for different EOSs. As can be seen from the graph, our result for the ideal EOS exactly matches with the Born rate so that the ratio is unity and it is independent of temperature. For the other two EOSs, we find significant enhancement of the DPR with respect to the Born rate. The enhancement is more at a higher temperature and as we increase $T$, we move towards the unity. Even at temperature $T=$ 250 MeV, we find $\simeq 15\%$ enhancement of the DPR with respect to the Born rate. At lower temperature, the enhancement will be even more.

\begin{figure}[h]
	\begin{center}
		\includegraphics[angle=-90, scale=0.35]{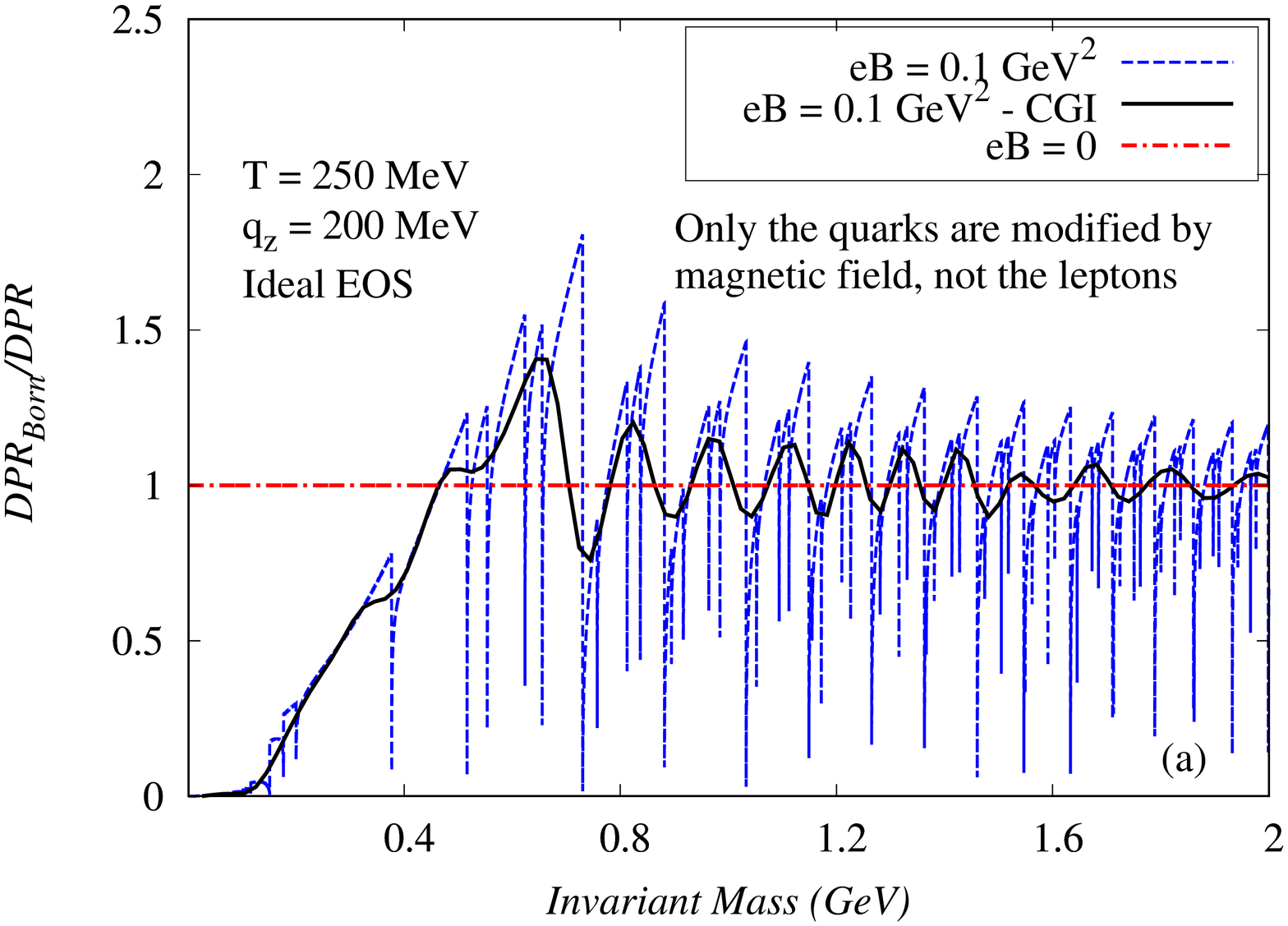}
		\includegraphics[angle=-90, scale=0.35]{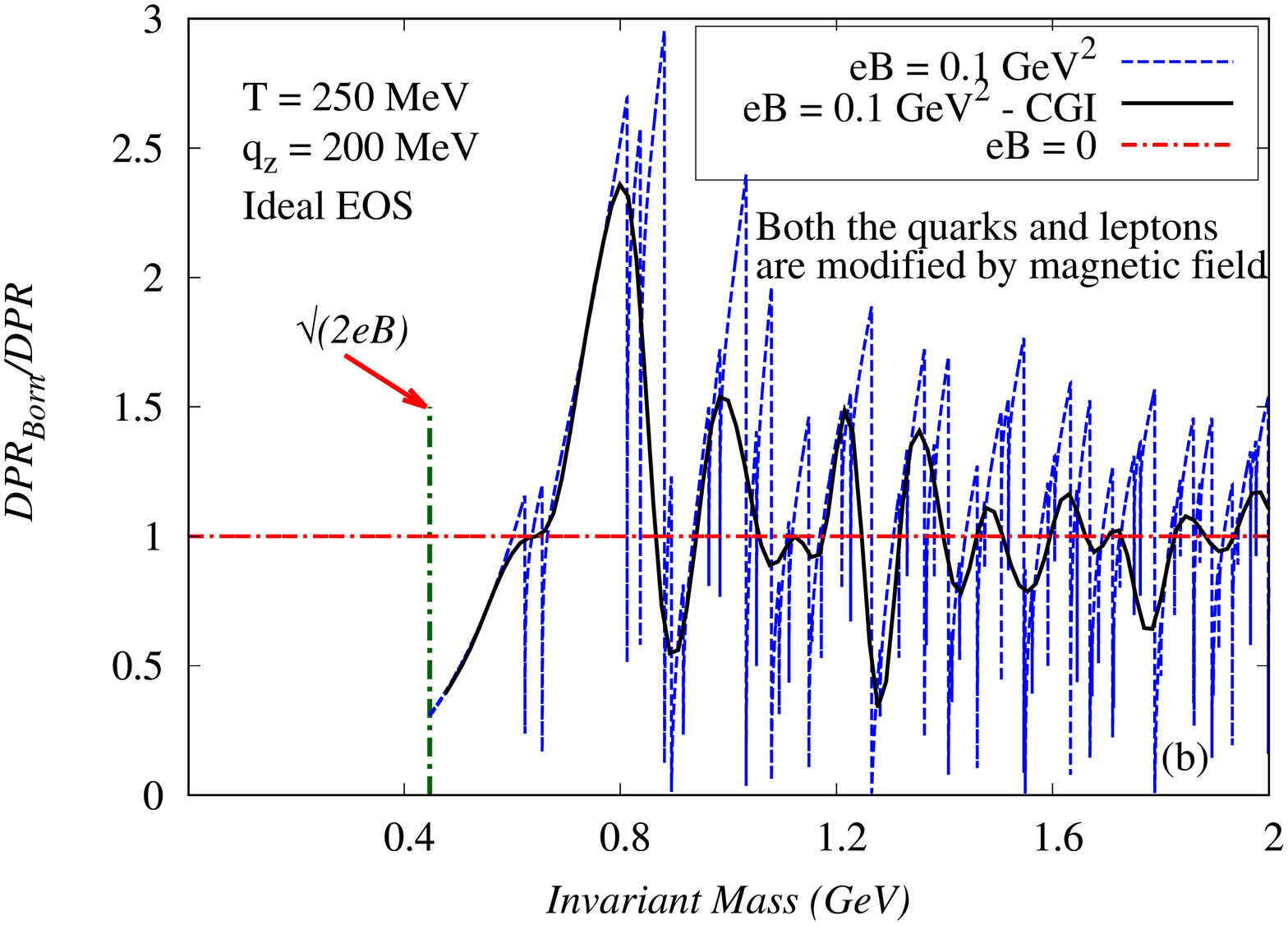}
	\end{center}
\caption{The ratio of Born rate to DPR as a function of dilepton invariant mass at $T=$ 250 MeV, $q_z$ = 200 MeV and at two different values of $eB$ (0 and 0.1 GeV$^2$) for Ideal EOS. Coarse-grained interpolated results are also shown for the non-zero $eB$ case. (a) Case where the quarks are modified by the magnetic field not the leptons and (b) case when both the quarks as well as lepton are modified by the magnetic field. The vertical green line in sub-figure (b) correspond to the DPR threshold $\sqrt{q_\parallel^2}=\sqrt{2eB}$.}
	\label{fig.dlr.2}
\end{figure}
\begin{figure}[h]
	\begin{center}
		\includegraphics[angle=-90, scale=0.35]{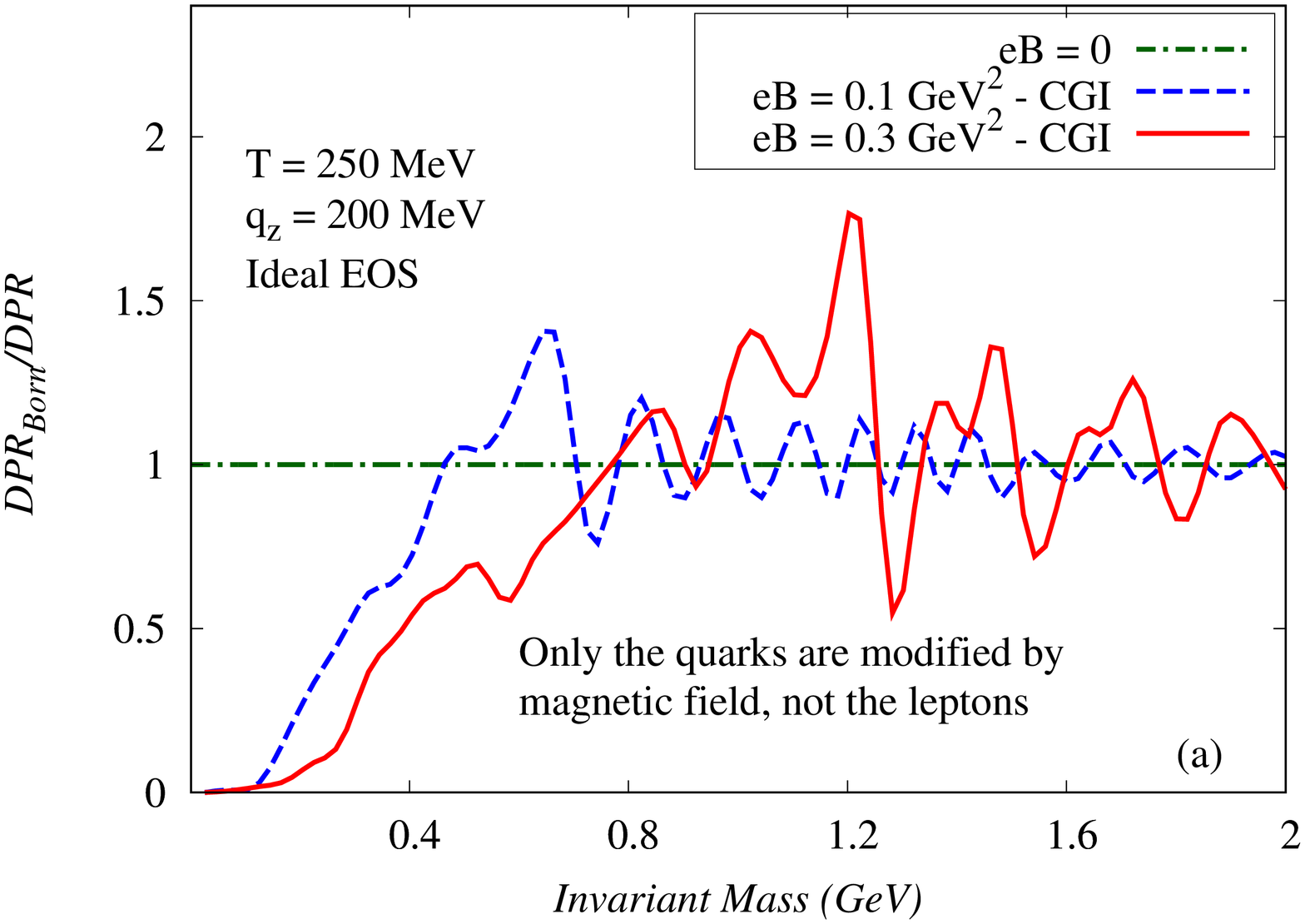}
		\includegraphics[angle=-90, scale=0.35]{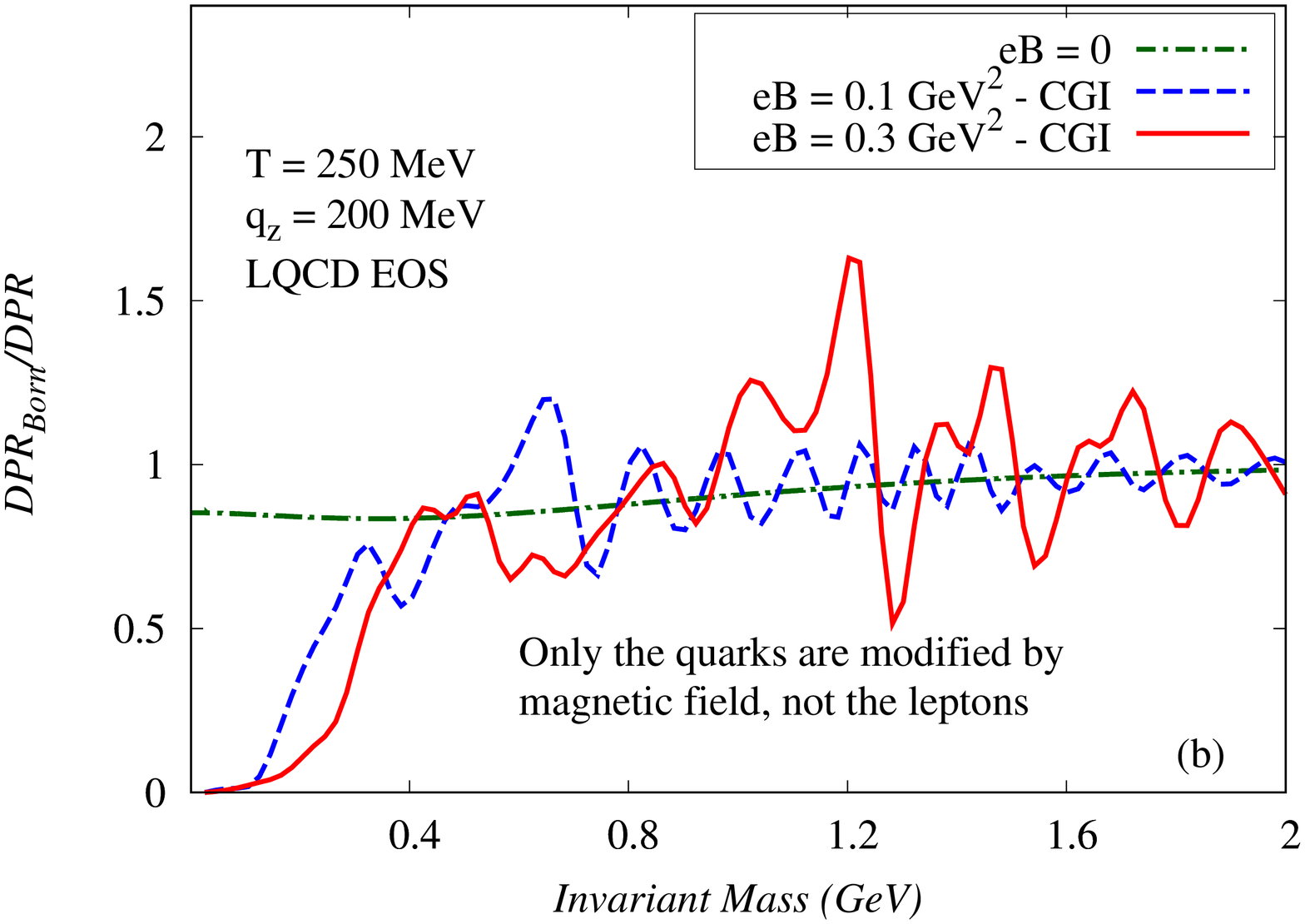} \\
		\includegraphics[angle=-90, scale=0.35]{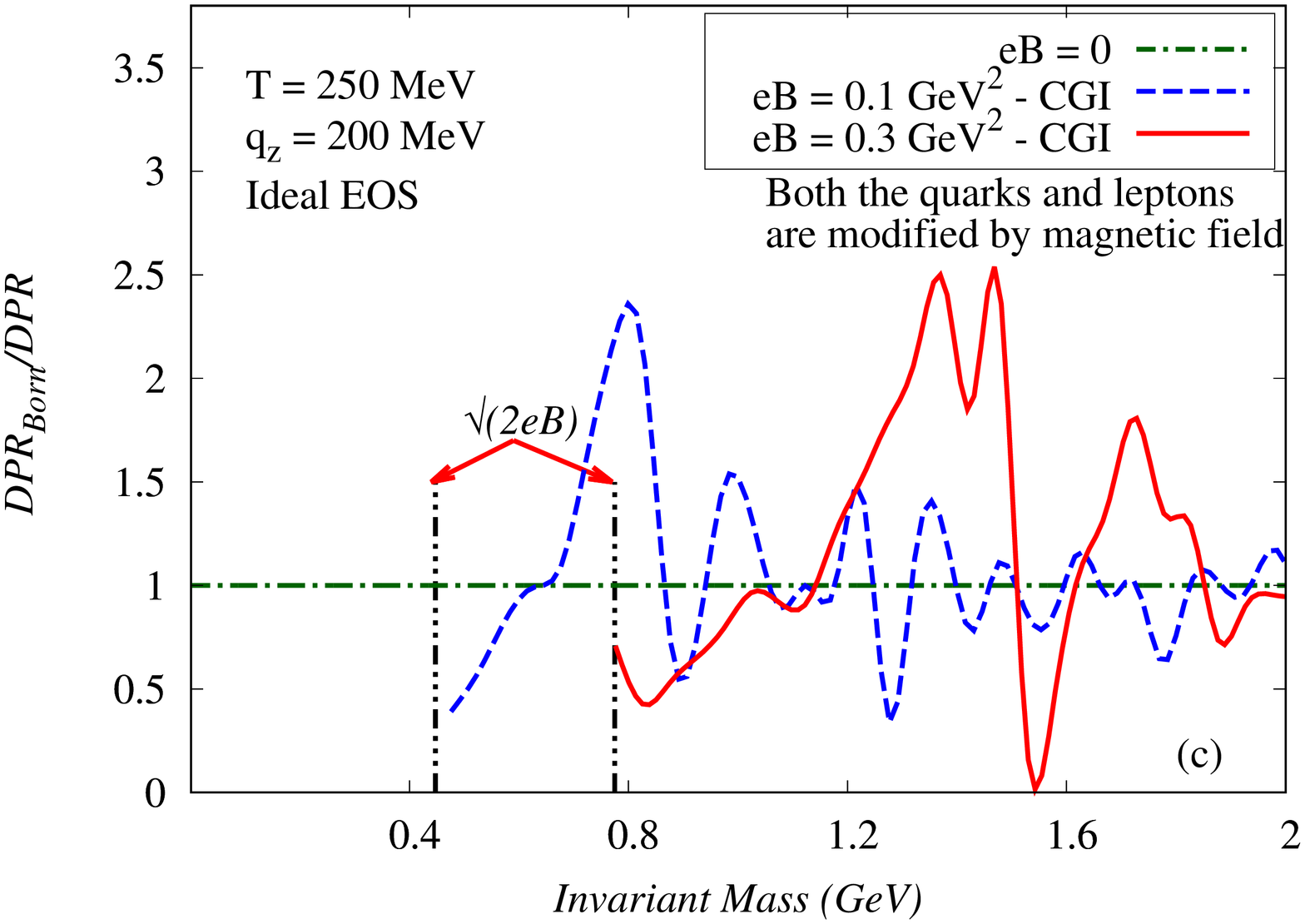}
		\includegraphics[angle=-90, scale=0.35]{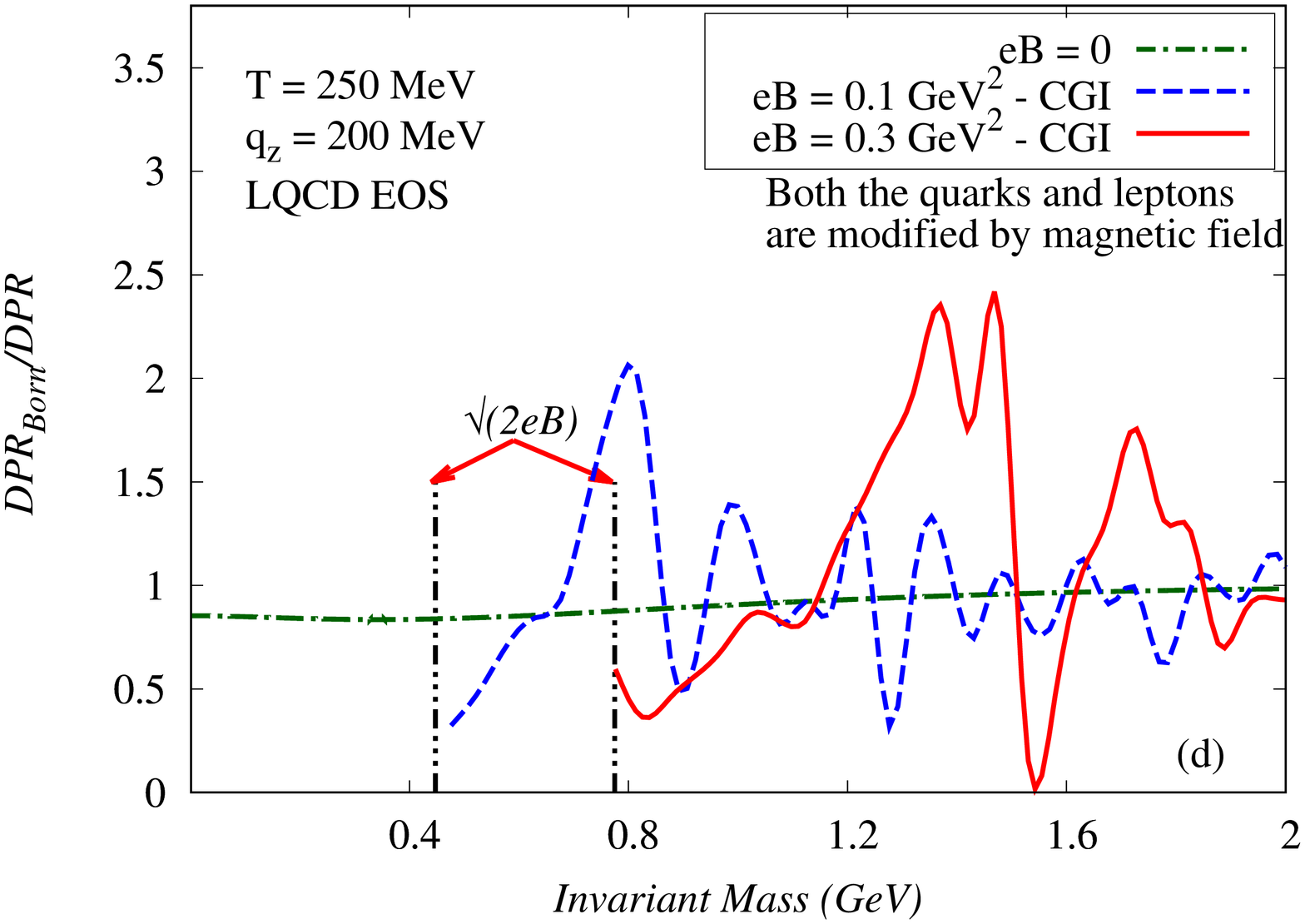} \\
	\end{center}
\caption{The ratio of Born rate to coarse-grained interpolated DPR as a function of dilepton invariant mass at $T=$ 250 MeV, $q_z$ = 200 MeV and at three different values of $eB$ (0, 0.1 and 0.3 GeV$^2$) for (left panel) ideal and (right panel) LQCD EOSs. The \textit{top panel} shows the case where the quarks are modified by the magnetic field not the leptons and \textit{bottom panel} shows the case when both the quarks as well as lepton are modified by the magnetic field. The vertical green lines in bottom panel correspond to the DPR threshold $\sqrt{q_\parallel^2}=\sqrt{2eB}$.}
	\label{fig.dlr.3}
\end{figure}

In presence of \textit{non-zero external magnetic}, the DPR is analyzed for the following two cases separately, {\it viz.}, (1) Only the quarks are affected by the external magnetic field not the dileptons and (2) Both the quarks as well as the dileptons are affected by the external magnetic field.

Now for, consistency check, we have first considered the LLL approximation and observe that the results of Ref.~\cite{Bandyopadhyay:2016fyd} could be reproduced by taking Ideal EOS and finite masses of quark flavours. In LLL approximation, the DPR vanishes for massless quark flavours as evident from Eq.~(\ref{eq.N.nl}) and also shown in Ref.~\cite{Bandyopadhyay:2016fyd}.

In Fig~\ref{fig.dlr.2}(a) and (b), we have plotted DPR$_\text{Born}$/DPR as a function of dilepton invariant mass at $eB=0.1$ GeV$^2$, $q_z=200$ MeV, $T=250$ MeV with ideal EOS for case-(1) and case-(2) respectively. The $eB=0$ graph is also shown for comparison. For both the case, we find significant enhancement of low invariant mass dilepton production which is due to the appearance of the Landau cuts in presence of $eB$ which was absent in $eB=0$ case. However at the higher invariant mass region, the DPR shows oscillatory behaviour about $eB=0$ graph. The oscillation amplitude is more for the case-(2). It is to be noted that, in case-(2), there is no dilepton production at $\sqrt{q_\parallel^2}<\sqrt{2eB}$ which is due to the threshold of $\rho^{l\bar{l}}$ as shown in Fig~\ref{fig.spectra.7}.

We conclude this section by finally presenting the CGI DPR$_\text{Born}$/DPR as a function of invariant mass of the dileptons for three different values of $eB$ (0, 0.1 and 0.3 GeV$^2$) in Fig.~\ref{fig.dlr.3}. Results are obtained at $T=250$ MeV and $q_z=200$ MeV. The left panel corresponds to ideal EOS whereas the right panel is for LQCD EOS. The top panel corresponds to the results for case-(1) whereas bottom panel depict the corresponding results for case-(2). With the increase in $eB$, the DPR enhances in the low invariant mass region for both the cases. However, due to kinematic threshold of the dilepton production at finite $eB$, the dilepton production starts from the invariant mass $\sqrt{q_\parallel^2}>\sqrt{2eB}$. At higher invariant mass regions, the ratio oscillates about the $eB=0$ graph and thus we get both the enhancement and suppression of dilepton production at different invariant mass.

\section{Summary and Conclusions} \label{sec.summary}

In summary, we have studied the electromagnetic spectral function and dilepton production rate in presence of both finite temperature as well as external magnetic field in a QCD medium. The photon polarization tensor at finite temperature is calculated using the RTF of TFT where the thermal distribution functions of loop particles (quarks/antiquarks) are modified in terms of their effective fugacities. The effective fugacity in the (quasi) quark distribution function encodes the hot QCD medium effect and is consistent with the realistic QCD EOSs like LQCD EOS. The effect of external magnetic field is introduced through the modification of the quark propagator in terms of Schwinger proper time propagator including all the Landau levels without any approximation on the strength of the external magnetic field. The Debye screening mass that has been obtained from the real part of the thermal photon self energy under external magnetic field turns out to be exactly the same as that could be obtained from linearized semi-classical transport theory. The imaginary part of the thermal photon self energy function has been obtained across the Unitary and Landau cuts. The Unitary cuts are already present in case of zero external magnetic field whereas the Landau cuts appeared purely due to the external magnetic field. Thereafter the analytical structure of the spectral function is studied and the magnetic field dependent thresholds of the Unitary and Landau cuts are obtained which are different from the $eB=0$ case. While calculating the DPR from thermal QGP medium, we have expressed it in terms of product of electromagnetic spectral functions due to quark and lepton loop. This formalism allows us to introduce the external magnetic field in terms of the modification of quark and lepton propagator by the Schwinger proper time one so that the knowledge of spin sum over leptonic spinor in presence of magnetic field is not required. We have observed spike like structure in the both electromagnetic spectral functions due to quark and lepton loop which arise because of the ``Threshold Singularity" in each Landau levels. The numerical results in the limit $eB\rightarrow0$ exactly reproduces the $eB=0$ result when the Ehrenfest's coarse-graining technique is used which smear out the threshold singularities.

In conclusion, the significant enhancement of the electromagnetic Debye screening mass with the increase in both the magnetic field and temperature has been observed, whereas, the inclusion of hot QCD medium effects decreases the magnitude of the Debye mass. The electromagnetic spectral function is seen to decrease with the increase in temperature and it has also non-trivial dependence on external magnetic. The spectral function in the low invariant mass region (dominated by the Landau term) increases with the increase in $eB$ whereas the high invariant mass region (dominated by the Unitary term) shows an oscillatory behaviour about the $eB=0$ values. The oscillation frequency (amplitude) decreases (increases) with the increase in $eB$. The spectral function due to lepton loop shows identical behaviour except the fact that in this case, the Landau cut does not appear. Thus the spectral function for lepton loop containing only the Unitary term has particular threshold in the invariant mass ($>\sqrt{2eB}$). This in turn restricts the kinematics for the low invariant mass dilepton production. A significant enhancement of the low invariant mass dilepton yield with respect to the Born rate has been observed due to the appearance of the Landau terms. On the other hand, in the high invariant mass region, a oscillation about the Born rate is noticed where the dilepton yield may enhances or suppresses depending on the invariant mass of the dilepton. Finally, our results are consistent with the predictions of other approaches (in limiting cases, whenever possible/applicable).

The immediate further extension of the work could be the inclusion of dissipative effects in hot medium and couple the analysis with (magneto) hydrodynamic framework in order to relate the relevant quantities to the experimentally realizable observables.

\section*{Acknowledgements}

S.G. would like to acknowledge Prof. Sourav Sarkar for the help in the field theoretical calculations. V.C. would like to acknowledge SERB, Govt. of India for Early Career Research Award (ECRA/2016/000683) and INSA-Department of Science and Technology , Govt. of India for INSPIRE Faculty award (IFA-13/PH-55). S.G. also acknowledges the Indian Institute of Technology Gandhinagar for the post doctoral fellowship. We are highly grateful to the people of India for their generous support for the research in fundamental sciences.

\appendix

\section{Useful Identities} \label{app.identities}
Using the orthogonality properties of the generalized Laguerre polynomials, one can derive the following identities
\begin{eqnarray}
\int\frac{d^2k_\perp}{(2\pi)^2}e^{-2\alpha_k}k_\perp^2L^1_{l-1}(2\alpha_k)L^1_{n-1}(2\alpha_k) 
&=& -\frac{(e_fB)^2}{16\pi}n\delta_{n-1}^{l-1} \\
\int\frac{d^2k_\perp}{(2\pi)^2}e^{-2\alpha_k}L_l(2\alpha_k)L_n(2\alpha_k) 
&=& \frac{|e_fB|}{8\pi}\delta_{n}^{l}
\end{eqnarray}
where, $\alpha_k=-k_\perp^2/|e_fB|$.


\section{Calculation of $\RE\overline{\Pi}^{00}(q^0=0,\vec{q}\rightarrow\vec{0})$ for $B\ne0$} \label{app.repi00}
In this appendix, we will sketch, how to obtain Eq.~(\ref{eq.md.b}). For this first we put $q_\perp=0$ in Eq.~(\ref{eq.re.pibar.tb}) and consider the $00$ component
\begin{eqnarray}
\RE\overline{\Pi}^{00}(q_\parallel) &=& \sum_{f}^{}e_f^2N_c
\sum_{l=0}^{\infty} \sum_{n=0}^{\infty}\bigintsss\frac{d^3k}{(2\pi)^3}\mathcal{P}
\TB{\frac{f(\omega_{k,l}^f)}{2\omega_{k,l}^f}\SB{\frac{N_{f,nl}^{00}(k^0=-\omega_{k,l}^f)}{(q^0-\omega_{k,l}^f)^2-(\omega_{p,n}^f)^2}
		+ \frac{N_{f,nl}^{00}(k^0=\omega_{k,l}^f)}{(q^0+\omega_{k,l}^f)^2-(\omega_{p,n}^f)^2}} \right. \nn \\
	&& \hspace{-1.0cm} \left. +
	\frac{f(\omega_{p,n}^f)}{2\omega_{p,n}^f}\SB{\frac{N_{f,nl}^{00}(k^0=-q^0-\omega_{p,n}^f)}{(q^0+\omega_{p,n}^f)^2-(\omega_{k,l}^f)^2}
		+ \frac{N_{f,nl}^{00}(k^0=-q^0+\omega_{p,n}^f)}{(q^0-\omega_{p,n}^f)^2-(\omega_{k,l}^f)^2}}} 
+ \RE\Pi_\text{vac}^{00}(q_\parallel) +\RE\Pi_\text{B}^{00}(q_\parallel,B)~.	\label{eq.repi.00}
\end{eqnarray} 
Calculating the trace over Dirac matrices in Eq.~(\ref{eq.Nnl}), we get,
\begin{eqnarray}
N_{f,nl}^{00} &=& (-1)^{n+l}e^{-2\alpha_k} 8 \TB{\frac{}{}-8L_{l-1}^1(2\alpha_k)L_{n-1}^1(2\alpha_k)k_\perp^2
+ \SB{\frac{}{}L_{l}(2\alpha_k)L_{n}(2\alpha_k)+L_{l-1}(2\alpha_k)L_{n-1}(2\alpha_k)} \times 
\right. \nn \\ && \hspace{3cm}\left.
\FB{m_f^2+k_0^2+k_z^2+k^0q^0+k_zq_z} \frac{}{}}~.
\end{eqnarray}
Substituting the above equation into Eq.~(\ref{eq.repi.00}) and performing the $d^2k_\perp$ integral using the identities given in Appendix~\ref{app.identities}, we get after putting $q^0=0$
\begin{eqnarray}
\RE\overline{\Pi}^{00}(q_z) &=& \sum_{f}^{}e_f^2N_c
\sum_{l=0}^{\infty} \sum_{n=0}^{\infty}\bigintsss_{-\infty}^\infty\frac{dk_z}{2\pi}\mathcal{P}
\TB{\frac{f(\omega_{k,l}^f)}{2\omega_{k,l}^f}\SB{\frac{\tilde{N}^{00}_{f,nl}(k^0=-\omega_{k,l}^f)
	+ \tilde{N}_{f,nl}^{00}(k^0=\omega_{k,l}^f)	
	}{(\omega_{k,l}^f)^2-(\omega_{p,n}^f)^2}
		} \right. \nn \\
	&& \hspace{-1.0cm} \left. +
	\frac{f(\omega_{p,n}^f)}{2\omega_{p,n}^f}\SB{\frac{\tilde{N}_{f,nl}^{00}(k^0=-\omega_{p,n}^f)+
		\tilde{N}_{f,nl}^{00}(k^0=\omega_{p,n}^f)}{(\omega_{p,n}^f)^2-(\omega_{k,l}^f)^2}
		 }} 
+ \RE\Pi_\text{vac}^{00}(q_z) +\RE\Pi_\text{B}^{00}(q_z,B)~.	\label{eq.repi.00.2}
\end{eqnarray} 
where, 
\begin{eqnarray}
\tilde{N}_{f,nl}^{00} &=& \FB{\frac{|e_fB|}{\pi}}\TB{\frac{}{}4|e_fB|n\delta_{n-1}^{l-1}
	+\FB{\delta_n^l+\delta_{n-1}^{l-1}}\FB{m_f^2+k_0^2+k_z^2+k_zq_z}}.
\end{eqnarray} 
The presence of the Kronecker delta function in the above equation will make the double sum of Eq.~(\ref{eq.repi.00.2}) into a single sum. Finally taking $q_z\rightarrow0$ limit of Eq.~(\ref{eq.repi.00.2}) we arrive at Eq.~(\ref{eq.re.pibar.tb}). It is worth mentioning that the last two terms of Eq.~(\ref{eq.repi.00.2}) goes to zero as $q_z\rightarrow 0$ so that they does not contribute in the Debye mass.


\section{Simplification of the Spectral function for $B\ne0$ case} \label{app.simplification.rho}
In view of simplifications in analytic calculations, we choose the transverse momentum of the photon $q_\perp=0$ so that for $B\ne0$, the spectral function $\rho(q^0,q_z)$ becomes, (using Eqs.~(\ref{eq.im.pibar.tb}) and (\ref{eq.rho.def}))
\begin{eqnarray}
\rho^{q\bar{q}}(q^0,q_z) &=& \frac{1}{4\pi\alpha}\sign{q^0}\tanh\FB{\frac{\beta q^0}{2}}\sum_{f}^{}e_f^2N_c
\sum_{l=0}^{\infty} \sum_{n=0}^{\infty}\pi\bigintsss\frac{d^3k}{(2\pi)^3}\frac{1}{4\omega_{k,l}^f\omega_{p,n}^f}
\TB{U^f_{1,nl}\delta\FB{q^0-\omega_{k,l}^f-\omega_{p,n}^f} \right. \nn \\ && \left.
	+U^f_{2,nl}\delta\FB{q^0+\omega_{k,l}^f+\omega_{p,n}^f} +L^f_{1,nl}\delta\FB{q^0+\omega_{k,l}^f-\omega_{p,n}^f}
	+ L^f_{2,nl}\delta\FB{q^0-\omega_{k,l}^f+\omega_{p,n}^f}}
\label{eq.rho.tb}
\end{eqnarray}
where,
\begin{eqnarray}
U^f_{1,nl} &=& \SB{1-f(\omega_{k,l}^f)-f(\omega_{p,n}^f)+2f(\omega_{k,l}^f)f(\omega_{p,n}^f)}
g_\munu N_{f,nl}^\munu(k^0=-\omega_{k,l}^f) \\
U^f_{2,nl} &=& \SB{1-f(\omega_{k,l}^f)-f(\omega_{p,n}^f)+2f(\omega_{k,l}^f)f(\omega_{p,n}^f)}
g_\munu N_{f,nl}^\munu(k^0=\omega_{k,l}^f) \\
L^f_{1,nl} &=& \SB{-f(\omega_{k,l}^f)-f(\omega_{p,n}^f)+2f(\omega_{k,l}^f)f(\omega_{p,n}^f)}
g_\munu N_{f,nl}^\munu(k^0=\omega_{k,l}^f) \\
L^f_{2,nl} &=& \SB{-f(\omega_{k,l}^f)-f(\omega_{p,n}^f)+2f(\omega_{k,l}^f)f(\omega_{p,n}^f)}
g_\munu N_{f,nl}^\munu(k^0=-\omega_{k,l}^f)
\end{eqnarray}
in which 
\begin{eqnarray}
N_{f,nl}^\munu(q_\parallel,k)g_\munu&=& (-1)^{n+l}e^{-\alpha_k-\alpha_p} 16 \TB{\frac{}{}-8L_{l-1}^1(2\alpha_k)L_{n-1}^1(2\alpha_p)(k_\perp^2+k_\perp\cdot q_\perp)
	\right. \nn \\ && \left.
	+ \SB{\frac{}{}L_{l}(2\alpha_k)L_{n}(2\alpha_p)+L_{l-1}(2\alpha_k)L_{n-1}(2\alpha_p)}m_f^2 \right. \nn \\ && \left.
	- \SB{\frac{}{}L_{l}(2\alpha_k)L_{n-1}(2\alpha_p)+L_{l-1}(2\alpha_k)L_{n}(2\alpha_p)}
	\FB{m_f^2-k_\parallel^2-k_\parallel\cdot q_\parallel}}~.
\end{eqnarray}
It is now trivial to perform the $d^2k_\perp$ integral in Eq.~(\ref{eq.rho.tb}) using the orthogonality of generalized Laguerre polynomials ( identities provided in Appendix~\ref{app.identities}) so that 
the spectral function becomes, 
\begin{eqnarray}
\rho^{q\bar{q}}(q^0,q_z) &=& \frac{1}{4\pi\alpha}\sign{q^0}\tanh\FB{\frac{\beta q^0}{2}}\sum_{f}^{}e_f^2N_c
\sum_{n=0}^{\infty} ~\sum_{l=(n-1)}^{(n+1)}\pi\bigintsss_{-\infty}^{\infty}
\frac{dk_z}{2\pi}\frac{1}{4\omega_{k,l}^f\omega_{p,n}^f}
\TB{\tilde{U}^f_{1,nl}\delta\FB{q^0-\omega_{k,l}^f-\omega_{p,n}^f} \right. \nn \\ && \left.
	+\tilde{U}^f_{2,nl}\delta\FB{q^0+\omega_{k,l}^f+\omega_{p,n}^f}
	+\tilde{L}^f_{1,nl}\delta\FB{q^0+\omega_{k,l}^f-\omega_{p,n}^f}
	+ \tilde{L}^f_{2,nl}\delta\FB{q^0-\omega_{k,l}^f+\omega_{p,n}^f}}
\label{eq.rho.tb2}
\end{eqnarray}
where,
\begin{eqnarray}
\tilde{U}^f_{1,nl} &=& \SB{1-f(\omega_{k,l}^f)-f(\omega_{p,n}^f)+2f(\omega_{k,l}^f)f(\omega_{p,n}^f)}
N_{f,nl}(k^0=-\omega_{k,l}^f) \\
\tilde{U}^f_{2,nl} &=& \SB{1-f(\omega_{k,l}^f)-f(\omega_{p,n}^f)+2f(\omega_{k,l}^f)f(\omega_{p,n}^f)}
N_{f,nl}(k^0=\omega_{k,l}^f) \\
\tilde{L}^f_{1,nl} &=& \SB{-f(\omega_{k,l}^f)-f(\omega_{p,n}^f)+2f(\omega_{k,l}^f)f(\omega_{p,n}^f)}
N_{f,nl}(k^0=\omega_{k,l}^f) \\
\tilde{L}^f_{2,nl} &=& \SB{-f(\omega_{k,l}^f)-f(\omega_{p,n}^f)+2f(\omega_{k,l}^f)f(\omega_{p,n}^f)}
N_{f,nl}(k^0=-\omega_{k,l}^f)
\end{eqnarray}
in which 
\begin{eqnarray}
N_{f,nl}(q_\parallel,k_\parallel)&=& (-1)^{n+l} \FB{\frac{|e_fB|}{\pi}}2\TB{4|e_fB|n\delta_{n-1}^{l-1}
	+\FB{\delta_n^l+\delta_{n-1}^{l-1}}m_f^2 + \FB{\delta_{n-1}^l+\delta_n^{l-1}}
	\FB{k_\parallel^2+k_\parallel\cdot q_\parallel-m_f^2}}~. 
\label{eq.N.nl}
\end{eqnarray} 
It is to be understood that a Kronecker delta with -ve index is zero i.e. $\delta_l^l=0$ if $l<0$ which comes from the fact that $L^a_{-1}(z)=0$ in the Schwinger propagator. The presence of Kronecker delta functions in the above equation have made the infinite double sums of Eq.~(\ref{eq.rho.tb}) into a single one or in other words the sum over $l$ runs from $(n-1)$ to $(n+1)$.

\section{Kinematic Domains of Spectral Function}\label{app.kinematic.domain}
Let us first consider the \textit{zero magnetic field} case. The expression for imaginary part of the self energy, given in Eq.~(\ref{eq.im.pibar.t}) contains four Dirac delta functions and they will be non-vanishing in certain kinematic regions. To see this, let us consider $\delta(q^0\mp\omega_k^f\mp\omega_p^f) = \delta(q^0\mp E)$ and $\delta(q^0\mp\omega_k^f\pm\omega_p^f) = \delta(q^0\mp E')$ where $E=(\omega_k^f+\omega_p^f)$ and $E'=(\omega_k^f-\omega_p^f)$. Both the functions $E=E(q,\vec{k},\cos\theta)$ and $E'=E'(q,\vec{k},\cos\theta)$ are defined in the domains $0\le|\vec{k}|<\infty$ and $|\cos\theta|\le1$ so that their co-domains come out to be
\begin{eqnarray}
\sqrt{\vec{q}^2+4m_f^2}\le E < \infty ~~~ \text{and} ~~~ -|\vec{q}|\le E' \le |\vec{q}|~.
\end{eqnarray}
Therefore, $\delta(q^0-E)$, $\delta(q^0+E)$ and $\delta(q^0\mp E')$ are non-vanishing in the kinematic regions defined in terms of $\sqrt{\vec{q}^2+4m_f^2} \le q^0 < \infty$ , $-\infty < q^0 \le -\sqrt{\vec{q}^2+4m_f^2}$ and $|q^0| \le |\vec{q}|$ respectively.

Let us now turn on the \textit{external magnetic field}. Analogous to the zero magnetic field case, the simplified spectral function in  Eq.~(\ref{eq.rho.tb2}) contains four Dirac delta functions namely $\delta\FB{q^0\mp\omega_{k,l}^f\mp\omega_{p,n}^f} = \delta(q^0\mp E^f_{nl})$ and $\delta\FB{q^0\mp\omega_{k,l}^f\pm\omega_{p,n}^f} = \delta(q^0\mp E'^f_{nl})$ where $E^f_{nl}=\omega_{k,l}^f+\omega_{p,n}^f$ and $E'^f_{nl}=\omega_{k,l}^f-\omega_{p,n}^f$. Both the functions $E^f_{nl}=E^f_{nl}(q^0,q_z,k_z)$ and $E'^f_{nl}=E'^f_{nl}(q^0,q_z,k_z)$ are defined in the domain $-\infty<k_z<\infty$ and their co-domains are given by 
\begin{eqnarray}
\sqrt{q_z^2+(m_{f,l}+m_{f,n})^2}\le E < \infty ~~~ \text{and} ~~~ \min\FB{q_z,E'_\pm}\le E' \le \max\FB{q_z,E'_\pm}
\end{eqnarray}
where, $E'_\pm = \frac{m_{f,l}-m_{f,n}}{\MB{m_{f,l}\pm m_{f,n}}}\sqrt{q_z^2+(m_{f,l}\pm m_{f,n})^2}$. The above equations implies that, for a particular set $\{n,l\}$, $\delta\FB{q^0-\omega_{k,l}^f-\omega_{p,n}^f}$ and $\delta\FB{q^0+\omega_{k,l}^f+\omega_{p,n}^f}$ will be non-vanishing in the kinematic regions $\sqrt{q_z^2+(m_{f,l}+m_{f,n})^2} \le q^0 < \infty$ and $-\infty < q^0 \le -\sqrt{q_z^2+(m_{f,l}+m_{f,n})^2}$ respectively. Whereas for $\delta\FB{q^0-\omega_{k,l}^f+\omega_{p,n}^f}$ and $\delta\FB{q^0+\omega_{k,l}^f-\omega_{p,n}^f}$, the corresponding kinematic regions will be respectively $\min\FB{q_z,E'_\pm}\le q^0 \le \max\FB{q_z,E'_\pm}$ and $-\max\FB{q_z,E'_\pm}\le q^0 \le -\min\FB{q_z,E'_\pm}$. Therefore, in Eq.~(\ref{eq.rho.tb2}), when the index $n$ is summed over from $0$ to $\infty$, $\delta\FB{q^0-\omega_{k,l}^f-\omega_{p,n}^f}$ and $\delta\FB{q^0+\omega_{k,l}^f+\omega_{p,n}^f}$ will be non-vanishing at $\sqrt{q_z^2+4m_f^2} \le q^0 < \infty$ and $-\infty < q^0 \le \sqrt{q_z^2+4m_f^2}$ respectively (for $n=0,l=0$). Corresponding regions for the non-vanishing of $\delta\FB{q^0\mp\omega_{k,l}^f\pm\omega_{p,n}^f}$ will be $|q^0|\le \sqrt{q_z^2+\FB{m_f-\sqrt{m_f^2+2|e_fB|}}^2}$ (for $n=1,l=0$). Note that, the thresholds of the Unitary cuts arise when both the two quarks in the loop lie in LLL. In contrast, the Landau cut threshold arise when one of the quarks in the loop is at LLL and the other one is at the next to LLL.


\section{Calculation of $W^{11}_\munu(q)$}\label{appendix.w11}
We have the expression of $W^{11}_\munu(q)$ from Eq.~(\ref{eq.w11}) as
\begin{eqnarray}
W^{11}_\munu(q) = i\int d^4xe^{iq\cdot x}\ensembleaverage{\mathcal{T}J_\mu(x)J^\dagger_\nu(0)}~.
\end{eqnarray}
Substituting the quark/antiquark current $J^\mu(x)$ from Eq.~(\ref{eq.quark.current.0}) into the above equation, we get
\begin{eqnarray}
W^{11}_\munu(q) = i\int d^4xe^{iq\cdot x}\sum_{f}^{}e_f^2
\ensembleaverage{\mathcal{T}\bar{q}_f(x)\gamma_\mu q_f(x)\bar{q}_f(0)\gamma_\nu q_f(0)}~.
\end{eqnarray}
Applying Wick's Theorem~\cite{Peskin:1995ev}, we get after some simplifications,
\begin{eqnarray}
W^{11}_\munu(q) = -iN_c\sum_{f}^{}e_f^2\int d^4xe^{iq\cdot x}\Tr\TB{\gamma_\mu S_f^{11}(x)\gamma_\nu S_f^{11}(-x)}
\label{eq.w11.1}
\end{eqnarray}
where, $S_f^{11}(x)$ is the $11$-component of the real time thermal quark propagator in coordinate space and can be Fourier transformed as
\begin{eqnarray}
S_f^{11}(x) = \ensembleaverage{\mathcal{T}q_f(x)\bar{q}_f(0)} = \int\frac{d^4p}{(2\pi)^4}e^{-ip\cdot x}(-iS^{11}_f(p))
\end{eqnarray}
in which $S^{11}_f(p)$ is defined in Eq.~(\ref{eq.S11.1}). Substituting the quark propagator into Eq.~(\ref{eq.w11.1}), we get
\begin{eqnarray}
W^{11}_\munu(q) = -iN_c\sum_{f}^{}e_f^2\int d^4xe^{iq\cdot x}\int\frac{d^4k_1}{(2\pi)^4}\int\frac{d^4k_2}{(2\pi)^4}
e^{-ix\cdot(k_1-k_2)}(-)\Tr\TB{\gamma_\mu S_f^{11}(k_1)\gamma_\nu S_f^{11}(k_2)}~.
\label{eq.w11.2}
\end{eqnarray}
It is now trivial to perform the $d^4x$ integral giving a Dirac delta function $\delta^4(q-k_1+k_2)$ which in turn is used to perform one of the $d^4k$ integral. Finally we get,
\begin{eqnarray}
W^{11}_\munu(q) = N_c\sum_{f}^{}e_f^2 i\int\frac{d^4k}{(2\pi)^4}
\Tr\TB{\gamma_\mu S_f^{11}(q+k)\gamma_\nu S_f^{11}(k)}~.
\label{eq.w11.3}
\end{eqnarray}


\section{Calculation of $L_\munu(q)$} \label{appendix.L}
We have the expression for $L_\munu(q)$ from Eq.~(\ref{eq.L}) as
\begin{eqnarray}
L_\munu(q) = i\int d^4x e^{iq\cdot x}\expectationvalue{0}{\mathcal{T}j^\dagger_\nu(x)j_\mu(0)}{0}~.
\end{eqnarray}
Substituting the lepton current $j^\mu(x)$ from Eq.~(\ref{eq.lepton.current.0}) into the above equation, we get
\begin{eqnarray}
L_\munu(q) = i\int d^4xe^{iq\cdot x}e^2\expectationvalue{0}{\mathcal{T}\bar{\psi}(x)\gamma_\nu \psi(x)\bar{\psi}(0)\gamma_\mu \psi(0)}{0}~.
\end{eqnarray}
Applying Wick's Theorem~\cite{Peskin:1995ev}, we get after some simplifications,
\begin{eqnarray}
L_\munu(q) = -ie^2\int d^4xe^{iq\cdot x}\Tr\TB{\gamma_\nu S_l(x)\gamma_\mu S_l(-x)}
\label{eq.L.1}
\end{eqnarray}
where, $S_l(x)$ is vacuum lepton propagator in coordinate space and can be Fourier transformed as
\begin{eqnarray}
S_l(x) = \expectationvalue{0}{\mathcal{T}\psi(x)\bar{\psi}(0)}{0} = \int\frac{d^4p}{(2\pi)^4}e^{-ip\cdot x}(-iS_l(p))
\end{eqnarray}
in which $S_l(p)=\frac{-(\cancel{p}+m)}{p^2-m^2+i\epsilon}$ is the corresponding vacuum momentum space lepton propagator. Substituting the lepton propagator into Eq.~(\ref{eq.L.1}), 
we get
\begin{eqnarray}
L_\munu(q) = -ie^2\int d^4xe^{iq\cdot x}\int\frac{d^4k_1}{(2\pi)^4}\int\frac{d^4k_2}{(2\pi)^4}
e^{-ix\cdot(k_1-k_2)}(-)\Tr\TB{\gamma_\nu S_l(k_1)\gamma_\mu S_l(k_2)}~.
\label{eq.L.2}
\end{eqnarray}
It is now trivial to perform the $d^4x$ integral giving a Dirac delta function $\delta^4(q-k_1+k_2)$ which in turn is used to perform one of the $d^4k$ integral. Finally we get,
\begin{eqnarray}
L_\munu(q) = e^2 i\int\frac{d^4k}{(2\pi)^4}\Tr\TB{\gamma_\nu S_l(q+k)\gamma_\mu S_l(k)}~.
\label{eq.L.3}
\end{eqnarray}


\section{Calculation of $L^+_\munu$ in Two Different Approaches} \label{appendix.2Ls}
In this appendix we will show that, both Eq.~(\ref{eq.L+.1}) and Eq.~(\ref{eq.L+.ImL}) gives identical analytic expression of $L^+_\munu(q)$. To see this, let us start with Eq.~(\ref{eq.L+.1})
\begin{eqnarray}
L^+_\munu(q) &=& (2\pi)^4\sum_{s_+,s_-}^{}\int\frac{d^3p_+}{(2\pi)^32E_+}\int\frac{d^3p_-}{(2\pi)^32E_-}
\delta^4(q-p_+-p_-)\expectationvalue{l^+l^-}{j_\mu(0)}{0} \expectationvalue{0}{j^\dagger_\nu(0)}{l^+l^-}~.
\end{eqnarray}
We now substitute lepton current $j^\mu(x)$ from Eq.~(\ref{eq.lepton.current.0}) into the above equation to get,
\begin{eqnarray}
L^+_\munu(q) &=& (2\pi)^4e^2\sum_{s_+,s_-}^{}\int\frac{d^3p_+}{(2\pi)^32E_+}\int\frac{d^3p_-}{(2\pi)^32E_-}
\delta^4(q-p_+-p_-)\expectationvalue{l^+l^-}{\bar{\psi}(0)\gamma_\mu\psi(0)}{0} \expectationvalue{0}{\bar{\psi}(0)\gamma_\nu\psi(0)}{l^+l^-}~.
\label{eq.L.4}
\end{eqnarray}
In order to calculate the matrix elements in the above equation, we use the Fourier decompositions of the lepton fields in terms of creation/annihilation operators as~\cite{Mallik:2016anp}
\begin{eqnarray}
\psi(x) &=& \sum_{s}^{}\int\frac{d^3p}{(2\pi)^32E_p}
\TB{\frac{}{}u(s,p)e^{-ip\cdot x}a(s,p)+v(s,p)e^{ip\cdot x}b^\dagger(s,p)} \\
\bar{\psi}(x) &=& \sum_{s}^{}\int\frac{d^3p}{(2\pi)^32E_p}
\TB{\frac{}{}\bar{v}(s,p)e^{-ip\cdot x}b(s,p)+\bar{u}(s,p)e^{ip\cdot x}a^\dagger(s,p)}
\end{eqnarray}
so that the matrix elements comes out to be
\begin{eqnarray}
\expectationvalue{l^+l^-}{\bar{\psi}(0)\gamma_\mu\psi(0)}{0} &=& \bar{u}(s_-,p_-)\gamma_\mu v(s_+,p_+) \\
\expectationvalue{0}{\bar{\psi}(0)\gamma_\nu\psi(0)}{l^+l^-} &=& \TB{\bar{u}(s_-,p_-)\gamma_\nu v(s_+,p_+)}^*~.
\end{eqnarray}
Substituting the matrix elements into Eq.~(\ref{eq.L.4}), we get,
\begin{eqnarray}
L^+_\munu(q) &=& (2\pi)^4e^2\sum_{s_+,s_-}^{}\int\frac{d^3p_+}{(2\pi)^32E_+}\int\frac{d^3p_-}{(2\pi)^32E_-}
\delta^4(q-p_+-p_-)\bar{u}(s_-,p_-)\gamma_\mu v(s_+,p_+)\TB{\bar{u}(s_-,p_-)\gamma_\nu v(s_+,p_+)}^* ~.\nn
\end{eqnarray}
Performing the spin sum over the lepton spinors, we get
\begin{eqnarray}
L^+_\munu(q) &=& (2\pi)^4e^2\int\frac{d^3p_+}{(2\pi)^32E_+}\int\frac{d^3p_-}{(2\pi)^32E_-}
\delta^4(q-p_+-p_-)\Tr\TB{(\cancel{p}_-+m)\gamma_\mu(\cancel{p}_+-m)\gamma_\nu} ~.
\label{eq.L.5}
\end{eqnarray}
Calculating the trace over Dirac matrices, and performing the $d^3p_+$ integral using the Dirac delta function we get, 
\begin{eqnarray}
L^{+\munu}(q) = \left.8\pi e^2\int\frac{d^3p_-}{(2\pi)^3}\frac{1}{4E_+E_-}\delta(q^0-E_+-E_-)
\TB{-g^\munu(m^2-p_-^2+p_-\cdot q)+(q^\mu p_-^\nu+q^\nu p_-^\mu)-2p_-^\mu p_-^\nu}\right|_{\vec{p}_+=\vec{q}-\vec{p}_-}
\label{eq.L.6}
\end{eqnarray}

Let un now consider Eq.~(\ref{eq.L+.ImL}), which can be written using Eq.~(\ref{eq.pi.1}) as
\begin{eqnarray}
L^+_\munu(q) = -2\IM\pi^\munu(q)
\label{eq.L.10}
\end{eqnarray}
where, $\pi^\munu(q)$ is the one-loop vacuum self energy of photon for a $l^+l^-$ loop. We have already calculated the one loop photon self energy for $q\bar{q}$ loop in Sec.~\ref{sec.self.t}. Thus $\pi^\munu$ is obtained from Eq.~(\ref{eq.im.pibar.t}) by replacing $T\rightarrow0$, $N_c\rightarrow1$, $\sum_{f}^{}e_f^2\rightarrow e^2$ and $m_f\rightarrow m$ so that we get,
\begin{eqnarray}
\IM\pi^\munu = \left.e^2\pi\int\frac{d^3k}{(2\pi)^3}\frac{1}{4E_kE_p}\delta(q^0-E_k-E_p)N^\munu(k^0=-E_k)
\right|_{\vec{p}=\vec{q}+\vec{k}}
\label{eq.pi.2}
\end{eqnarray}
where, $N^\munu(q)$ can be read from Eq.~(\ref{eq.N.t}) as
\begin{eqnarray}
N^\munu(q,k) = \Tr\TB{\gamma^\nu\FB{\cancel{q}+\cancel{k}+m}\gamma^\mu\FB{\cancel{k}+m}} 
= 4\TB{(m^2-k^2-k\cdot q)g^\munu+2k^\mu k^\nu+(k^\mu q^\nu+k^\nu q^\mu)}~.
\end{eqnarray}
It is worth mentioning that we have considered only the physical kinematic region $q^2>0, q^0>0$ while writing Eq.~(\ref{eq.pi.2}) so that the Unitary-II cut does not contribute. It is now trivial to check that after a change of variable $\vec{k}\rightarrow-\vec{k}$ in Eq.~(\ref{eq.pi.2}) followed by substitution into Eq.~(\ref{eq.L.10}) leads to the same expression of $L^+_\munu(q)$ as obtained earlier in Eq.~(\ref{eq.L.6}).

\bibliographystyle{apsrev4-1}
\bibliography{snigdha}

\end{document}